\documentclass[aps,prd,reprint,amsmath,amssymb,showpacs,preprintnumbers,superscriptaddress,nofootinbib,longbibliography]{revtex4-1}

\usepackage{slashed}
\usepackage{bm}
\usepackage{latexsym,amssymb,amsmath,float,url,mathrsfs}
\usepackage{bbold}
\usepackage{latexsym}
\usepackage{graphicx}
\usepackage{epstopdf}
\usepackage{amsmath}
\usepackage{subfigure}
\usepackage{hyperref}
\usepackage[dvipsnames]{xcolor}

\hypersetup{
     colorlinks   = true,
     citecolor    = violet,
     urlcolor     = violet,
     linkcolor    = violet
}

\begin{document}

\title{Powerful Solar Signatures of Long-Lived Dark Mediators}

\author{Rebecca K.\ Leane}
\affiliation{Center for Cosmology and AstroParticle Physics (CCAPP), Ohio State University, Columbus, OH 43210, USA}
\affiliation{ARC Centre of Excellence for Particle Physics at the Terascale\\
School of Physics, The University of Melbourne, Victoria 3010, Australia}

\author{Kenny C.\ Y.\ Ng}
\affiliation{Center for Cosmology and AstroParticle Physics (CCAPP), Ohio State University, Columbus, OH 43210, USA}
\affiliation{Department of Physics, Ohio State University, Columbus, OH 43210, USA}
\affiliation{Department of Particle Physics and Astrophysics, Weizmann Institute of Science, Rehovot 76100, Israel}

\author{John F.\ Beacom} 
\affiliation{Center for Cosmology and AstroParticle Physics (CCAPP), Ohio State University, Columbus, OH 43210, USA} 
\affiliation{Department of Physics, Ohio State University, Columbus, OH 43210, USA}
\affiliation{Department of Astronomy, Ohio State University, Columbus, OH 43210, USA \\ 
{\tt \href{mailto:rebecca.leane@gmail.com}{rebecca.leane@gmail.com}, \href{mailto:chun-yu.ng@weizmann.ac.il}{chun-yu.ng@weizmann.ac.il}, \href{mailto:beacom.7@osu.edu}{beacom.7@osu.edu}} \\
{\tt \footnotesize \href{http://orcid.org/0000-0002-1287-8780}{0000-0002-1287-8780}, \href{http://orcid.org/0000-0001-8016-2170}{0000-0001-8016-2170}, \href{http://orcid.org/0000-0002-0005-2631}{0000-0002-0005-2631} \smallskip}}

\date{\today}

\begin{abstract}
Dark matter capture and annihilation in the Sun can produce detectable high-energy neutrinos, providing a probe of the dark matter-proton scattering cross section. 
We consider the case when annihilation proceeds via long-lived dark mediators, which allows gamma rays to escape the Sun and reduces the attenuation of neutrinos. For gamma rays, there are exciting new opportunities, due to detailed measurements of GeV solar gamma rays with Fermi, and unprecedented sensitivities in the TeV range with HAWC and LHAASO. For neutrinos, the enhanced flux, particularly at higher energies~($\sim$TeV), allows a more sensitive dark matter search with IceCube and KM3NeT. We show that these search channels can be extremely powerful, potentially improving sensitivity to the dark matter spin-dependent scattering cross section by several orders of magnitude relative to present searches for high-energy solar neutrinos, as well as direct detection experiments.
\end{abstract}

\maketitle

\section{Introduction}
\label{sec:intro}

There is overwhelming evidence that dark matter (DM) is the dominant form of matter in the universe~\cite{Bertone:2004pz}. However, across experimental tests of its annihilation, scattering, and production processes, no details of its fundamental nature have yet been revealed. For models with unsuppressed spin-independent scattering interactions, there are severe bounds on the properties of DM from direct detection experiments, such as LUX~\cite{Akerib:2013tjd,Akerib:2015rjg} and PandaX-II~\cite{Tan:2016zwf}. If instead there are only spin-dependent interactions, a much larger part of the parameter space remains uninvestigated, with best limits currently set by LUX~\cite{Akerib:2016lao} and PandaX-II~\cite{Fu:2016ega} for neutron scattering, and PICO-60 C$_3$F$_8$ \cite{Amole:2017dex} for proton scattering.

The Sun is an alternate probe, as it can gravitationally capture DM \cite{1985ApJ296679P,Krauss:1985ks,Silk:1985ax,Peter:2009mk}, after DM loses energy through scattering with solar nucleons. If DM is captured, it must have scattering interactions that force further energy loss and accumulation in the solar core, leading to annihilation to Standard Model (SM) particles. Measurement of these SM particles provides insight to the nature of DM. However, in order to escape the Sun for detection, the particles need to be very weakly interacting. Amongst the potential SM particles produced in the solar core, only neutrinos can escape. Even then, there is significant attenuation for neutrinos above about 100 GeV.

As DM has not yet been found, more general theoretical scenarios should be considered. A fairly minimal scenario consists of a DM candidate, along with a new particle to mediate interactions between the dark and visible sectors. An interesting possibility is that, as a consequence of particular model properties, the mediator may have a long decay lifetime. These `long-lived dark mediators' are well motivated, and include examples such as the dark photon~\cite{Kobzarev:1966qya, Okun:1982xi, HOLDOM1986196, HOLDOM198665}, dark Higgs~\cite{Batell:2009zp}, and many supersymmetric particles~\cite{Martin:1997ns}. There is also wide interest in searches at current~\cite{Reece:2009un,Morrissey:2014yma,Aad:2015rba,Eigen:2015rea,Lees:2015rxq} and future colliders~\cite{Alekhin:2015byh}.
\begin{figure*}
     \begin{center}
        \subfigure{
            \includegraphics[width=0.4912\textwidth]{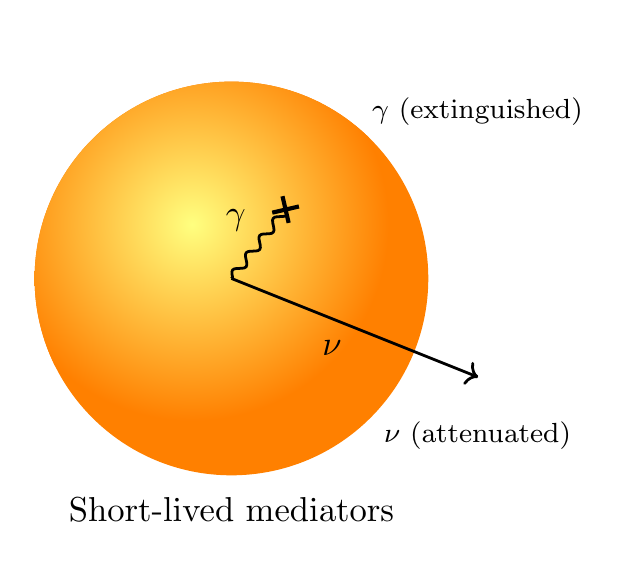}}
        \subfigure{
           \includegraphics[width=0.4912\textwidth]{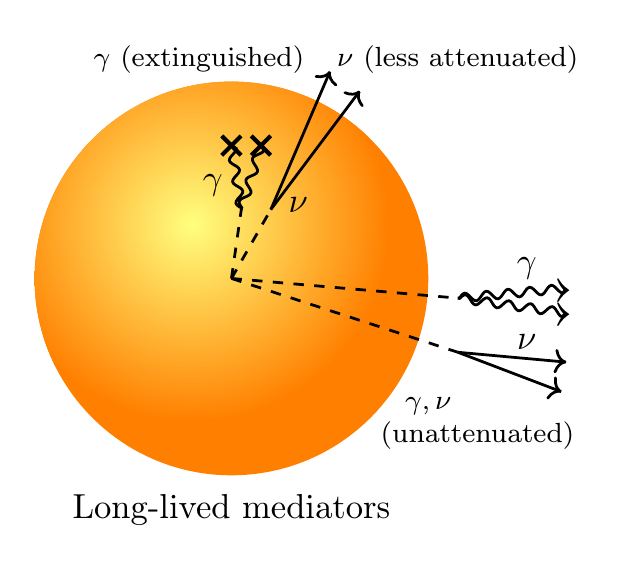}}
    \end{center}
    \caption{{\bf Left:} Short-lived mediator scenario (usual case). Only neutrinos can escape the Sun and they are attenuated. \\{\bf Right:} Long-lived dark mediator scenario. Gamma rays can escape, and neutrinos are less attenuated.}
   \label{fig:sun}
\end{figure*}

Figure~\ref{fig:sun} illustrates how the long-lived mediator setup can strongly affect solar DM detection: the mediator can decay outside of the solar core, producing otherwise attenuated or lost solar DM signals~\cite{Pospelov:2007mp,Pospelov:2008jd, Batell:2009zp,Rothstein:2009pm,Chen:2009ab,Schuster:2009au,Schuster:2009fc,Meade:2009mu,Bell:2011sn,Feng:2015hja,Kouvaris:2016ltf,Feng:2016ijc,Kouvaris:2016ltf,Adrian-Martinez:2016ujo,Allahverdi:2016fvl, Ardid:2017lry}. While it is known that prospects are improved in this scenario, investigations to date are not complete on considering the full range of data and models.

In this paper, we examine the prospects for new gamma-ray and neutrino experiments in a \mbox{model-independent} framework. For gamma rays, this is particularly pertinent with new detailed measurements of the Sun in the GeV range with the Fermi Large Area Telescope (Fermi-LAT)~\cite{2011ApJ734116A,Ng:2015gya}, as well as unprecedented sensitivity to TeV gamma rays with upcoming analyses from the High Altitude Water Cherenkov (HAWC) Observatory~\cite{Abeysekara:2013tza,Abeysekara:2017mjj}, which began operating in 2015, and the Large High Altitude Air Shower Observatory (LHAASO)~\cite{Zhen:2014zpa}, which is under construction and expected to begin operating in 2020. For neutrinos, this is particularly pertinent for the multi-TeV window at IceCube~\cite{Aartsen:2016exj}, and future neutrino telescopes such as KM3NeT~\cite{Adrian-Martinez:2016fdl}. We demonstrate these telescopes and observatories can provide DM probes orders of magnitude stronger than both current searches for high-energy solar neutrinos, and 
direct detection 
experiments.

We define the sensitivity to such scenarios in the following ways:  On the theory side, we consider optimal cases, for example where the mediators decay just outside the Sun.  On the experimental side, we are more conservative, requiring that the new signals be as large as measurements, not just their uncertainties. Accordingly, we aim for a precision of a factor of a few, neglecting some smaller effects. This optimal scenario will demonstrate the full power of long-lived mediators for solar DM searches.  Our sensitivity can be mapped to the parameter spaces of any particular model realizations, together with any other constraints, which will be a subset of the space considered.  Therefore, we focus on the new signatures and the experimental sensitivity. 

In Sec.~\ref{sec:solardetails}, we review the processes for DM capture and annihilation in the Sun. In Sec.~\ref{sec:mediator}, we discuss the modifications for the long-lived mediator scenario. We then demonstrate the power of gamma-ray signals with \mbox{Fermi-LAT}, HAWC, and LHAASO in Sec.~\ref{sec:limits_gamma}, and for neutrinos with IceCube and KM3NeT in Sec.~\ref{sec:limits_neutrino}. In Sec.~\ref{sec:models}, we discuss interpretations of our results in the context of popular models. Finally, other constraints are discussed in Sec.~\ref{sec:constraints} before concluding in Sec.~\ref{sec:conclusions}.

\section{Dark Matter Solar Capture and Annihilation}
\label{sec:solardetails}

The usual scenario for DM capture and annihilation in the Sun has been well studied \cite{1985ApJ296679P,Krauss:1985ks,Silk:1985ax,Press:1985ug,Silk:1985ax,Krauss:1985aaa,Griest:1986yu,Gould:1987ww,Gould:1987ir,Gould:1991hx,Batell:2009zp,Peter:2009mk,Feng:2016ijc}.
DM is gravitationally captured by the Sun if it loses sufficient energy after scattering with solar nuclei. As the captured DM accumulates in the Sun's core, there are more DM particles available to power DM annihilation. However, annihilation depletes the DM supplied by capture. Therefore, the total number of DM particles in the solar core is determined by an interplay of the capture rate $\Gamma_{\rm cap}$ and annihilation rate $\Gamma_{\rm ann}$. Equilibrium is reached if the equilibrium timescale is less than the age of the Sun. 

In the regime that DM self-interactions~\cite{Zentner:2009is} are not relevant, the relation of these processes and the number of DM particles $N_\chi$ in the Sun at time $t$ is given by
\begin{equation}
  \frac{d}{dt}N_\chi=\Gamma_{\rm cap}-C_{\rm ann}N_\chi^2\, ,
  \label{eq:numberrate}
 \end{equation}
where $\Gamma_{\rm cap}$ is the DM capture rate and $C_{\rm ann}$ is a coefficient that describes the annihilation processes.
The number of DM particles in the Sun rapidly approaches equilibrium when $t>t_{\rm equil}=1/\sqrt{\Gamma_{\rm cap}C_{\rm ann}}$.   
We focus on DM masses above 4 GeV, where evaporation is irrelevant~\cite{GRIEST1987681,PhysRevD.34.2206}.
Equilibrium thus depends on sufficiently large scattering and annihilation cross sections.
For the scenarios considered here, both conditions can be met \cite{Rott:2011fh}. Therefore, a simple relation between annihilation rate and capture rate is obtained,
\begin{equation}
  \Gamma_{\rm ann}= \frac{1}{2}C_{\rm ann}N_{\chi}^2 = \frac{1}{2}\Gamma_{\rm cap},
  \label{eq:ann}
 \end{equation}
independent of the DM annihilation cross section. \\ \\

\section{Long-Lived Dark Mediator Scenario}
\label{sec:mediator}

\subsection{Opportunities and Framework}

The energy flux of DM annihilation products in the Sun is enormous. For example, in the case that \mbox{100 GeV} or 1 TeV DM with spin-dependent scattering cross sections of $\sim10^{-40}{\rm \, cm^2}$ (capture rates of $10^{22}\,$s$^{-1}$ and $10^{20}\,$s$^{-1}$, respectively) annihilates directly to gamma rays, the energy fluxes are
\begin{subequations}
 \label{eq:gammaflux}
\begin{gather}
   E_\gamma^2 \frac{d\Phi_{\gamma}}{dE_\gamma}\sim 10^{-4} {\rm \,GeV\, cm^{-2}\,s^{-1}} \ \ \  \ m_\chi=100\,{\rm GeV},\\
   E_\gamma^2 \frac{d\Phi_{\gamma}}{dE_\gamma} \sim 10^{-5} {\rm \,GeV\, cm^{-2}\,s^{-1}}  \ \ \ \ \ \ \ m_\chi=1\,{\rm TeV},
\end{gather}
\end{subequations}
where we have assumed that the gamma-ray spectrum is measured in bins one decade wide  (this is $\sim$10 times too conservative for Fermi, but appropriate for HAWC).
The best experimental sensitivity to 100 GeV solar gamma rays comes from Fermi-LAT, with sensitivity
\begin{equation}
E_\gamma^2 \frac{d\Phi_{\gamma}}{dE_\gamma}\sim 10^{-8} {\rm \,GeV\, cm^{-2}\,s^{-1}},
 \label{eq:gammaflux}
\end{equation}
while the best sensitivity to 1 TeV gamma rays is from HAWC,
\begin{equation}
E_\gamma^2 \frac{d\Phi_{\gamma}}{dE_\gamma}\sim 10^{-9} {\rm \,GeV\, cm^{-2}\,s^{-1}}.
 \label{eq:gammaflux}
\end{equation}
In each case, the annihilation flux is in excess of sensitivity by a factor of $10^4$. Of course, in the usual scenario, the difficulty is that it is not possible to observe these promptly extinguished gamma rays. This is why the long-lived dark mediator scenario is so compelling --- 
gamma rays can escape the solar core, providing a probe of the immense annihilation flux.

For solar gamma rays~\cite{Batell:2009yf}, the sensitivity to long-lived mediators from these experiments has not yet been fully explored. For solar neutrinos, limits exist for short-lived mediators \cite{Aartsen:2016exj}, but the improvements from long-lived mediators through less absorption of neutrinos \cite{Bell:2011sn} have also not been fully quantified.

The energy flux from DM annihilation in the Sun is
\begin{equation}
 E^2 \frac{d\Phi}{dE}=\frac{\Gamma_{\rm ann}}{4\pi D_\oplus^2}\times E^2\frac{dN}{dE}\times {\rm Br(Y\rightarrow SM)}\times P_{\rm surv},
 \label{eq:flux}
\end{equation}
where $D_\oplus=1$ A.U. is the average distance between the Sun and the Earth, $E^2 dN/dE$ is the spectrum per DM annihilation, ${\rm Br(Y\rightarrow SM)}$ is the branching fraction of the mediator $Y$ to SM particles, and $P_{\rm surv}$ is the probability of the signal surviving to reach the detector, which includes factors such as attenuation and mediator decay length. While this factor diminishes the flux, the cost to the total flux pales in comparison to the net gain in exploiting the large annihilation flux in the Sun. In the standard scenario, $P_{\rm surv}=0$ for gamma rays, and is exponentially suppressed for neutrinos with energies above about 100~GeV.

In the following subsections, assumptions and properties of long-lived mediators relevant to each of the terms in Eq.~(\ref{eq:flux}) are described.

\subsection{Annihilation Rate}
After equilibrium is reached, the annihilation rate of DM in the Sun, $\Gamma_{\rm ann}$, is related to the capture rate as per Eq.~(\ref{eq:ann}).  We use {\sc DarkSUSY} \cite{Gondolo:2004sc} to compute the annihilation rate $\Gamma_{\rm ann}$ for a given DM scattering cross section and mass. The capture rate scales $\propto m_\chi^{-1}$ up to a few 10 GeV, which follows the local DM number density.  Above a few 100 GeV, it scales $\propto m_\chi^{-2}$, due to kinematic suppression of the energy loss~\cite{Jungman:1995df,Rott:2012qb,Choi:2013eda}.

\subsection{Branching Fractions}

We assume a $100\%$ branching fraction of the mediator to each final state in turn, which is the optimal case.  If only one final state produces observable signals, it is straightforward to scale our result with the branching fraction.  The effects of considering multiple final states in the context of specific models are discussed in Sec.~\ref{sec:models}.
 
\subsection{Energy Spectra} \label{sec:spectra}
\begin{figure*}
     \begin{center}
        \subfigure{
            \includegraphics[width=0.49\textwidth]{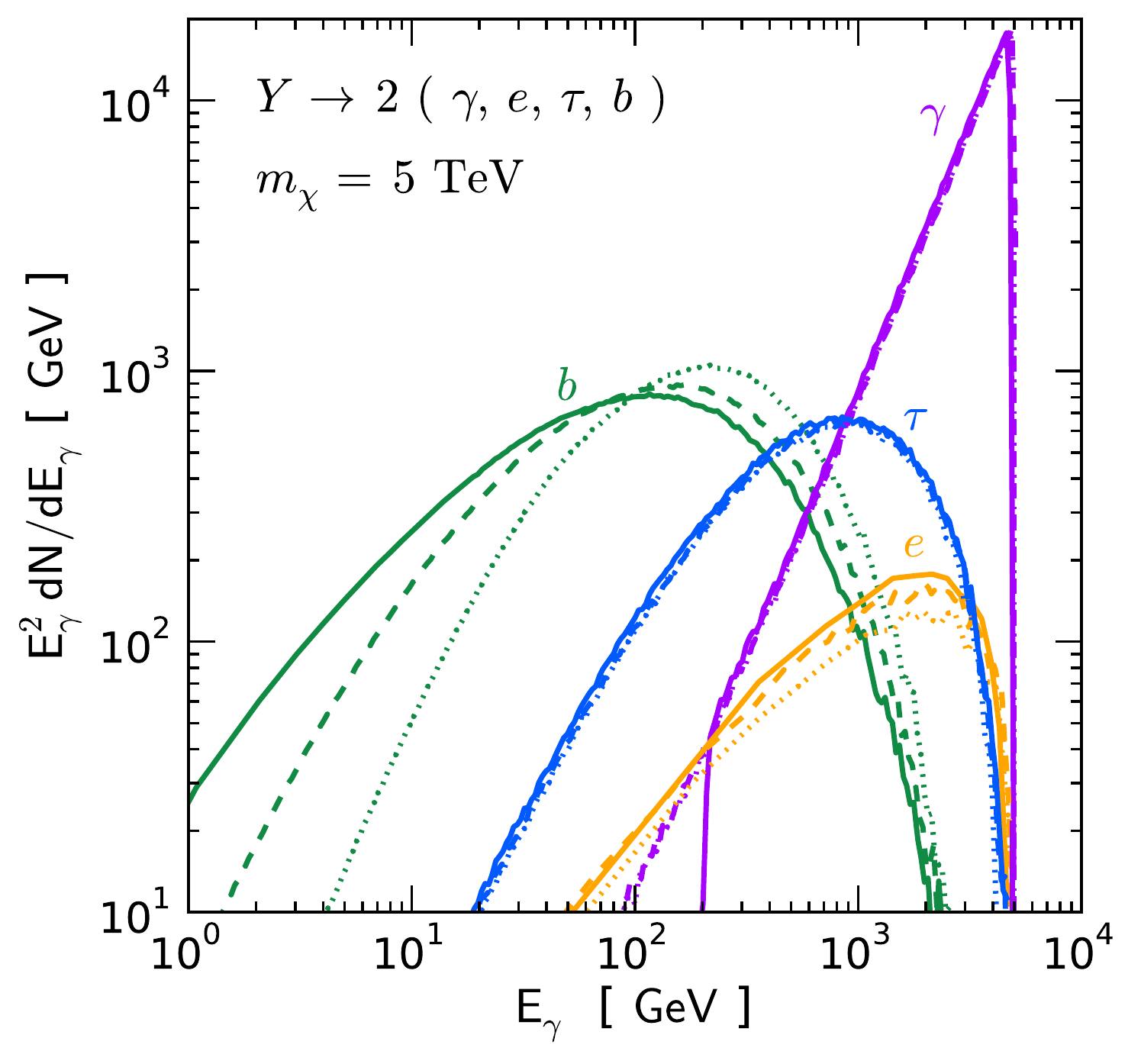}}
        \subfigure{
           \includegraphics[width=0.49\textwidth]{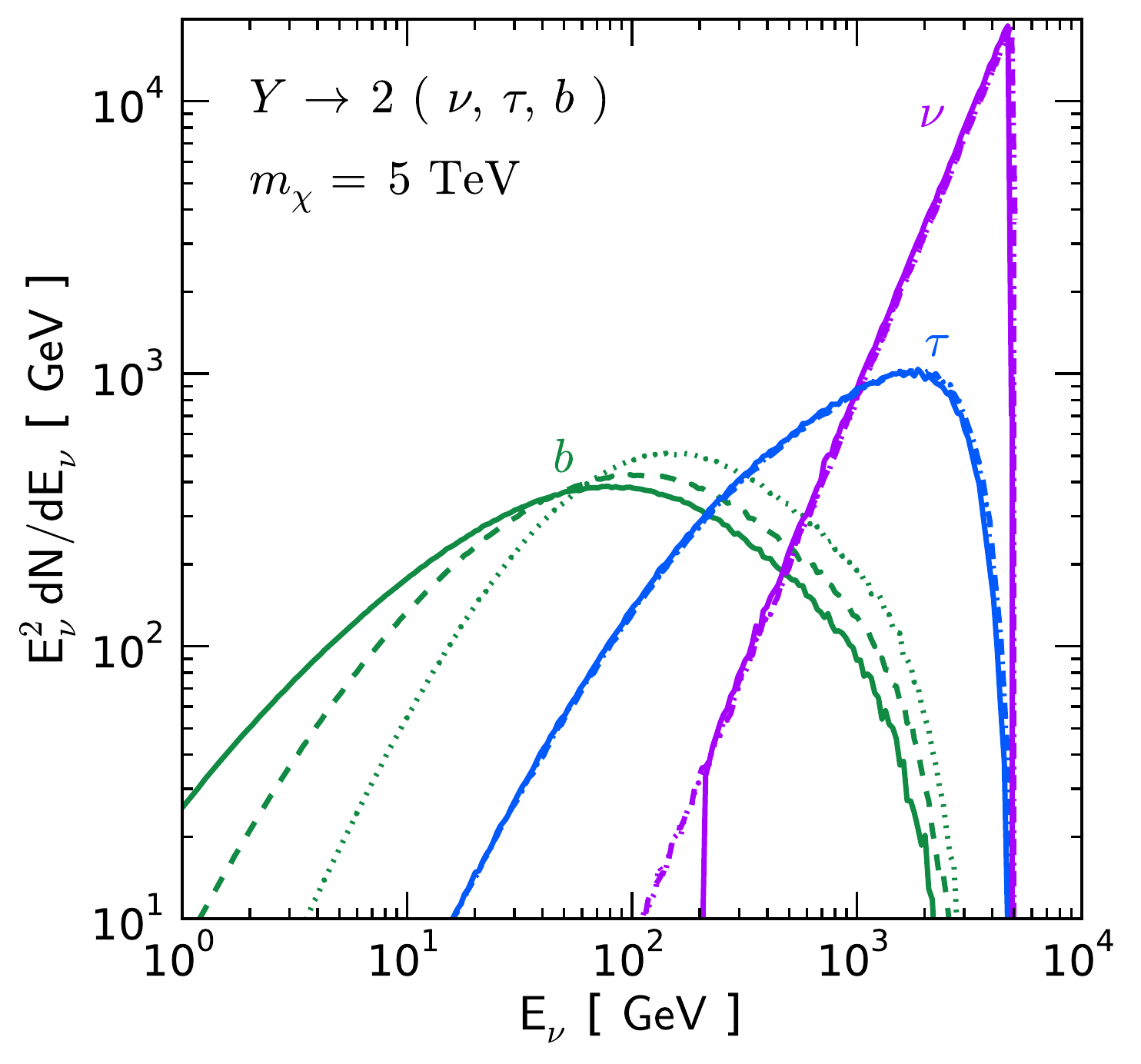}}
    \end{center}
\caption{{\bf Left:} Gamma-ray spectra $E_{\gamma}^2 dN/dE_{\gamma}$ for various final states, per DM annihilation, with mediator masses \mbox{$m_Y=2$~TeV (solid)}, $m_Y=200$ GeV (dashed), and $m_Y=20$ GeV (dotted). {\bf Right:} Neutrino spectra.}
\label{fig:mediator_compare2}
\end{figure*}
DM annihilates to long-lived mediators as 
\begin{equation}
 \chi\chi\rightarrow YY \rightarrow 2 \,(\,{\rm SM + \overline{SM}}\,) \rightarrow ... \gamma,\nu...
\end{equation}
where the mediator $Y$ decays to two SM particles, which consequently can decay into or radiate gamma rays or neutrinos.

We use {\sc Pythia} \cite{Sjostrand:2014zea} to generate the neutrino and gamma-ray energy spectra, where an effective resonance with energy $2m_\chi$ decays to two mediators. Depending on the final state particles, gamma rays or neutrinos can arise from direct decay to $4\gamma$ or $4\nu$, electroweak bremsstrahlung, or consequent particle decays. Our simulations take into account all these possibilities where relevant, and are the fully decayed spectra in vacuum. 

Figure~\ref{fig:mediator_compare2} shows that the energy spectra from DM annihilation are approximately the same for processes that are topologically identical. That is, for a given DM mass to a given $n$-body final state, approximately the same energy spectra is obtained regardless of the fundamental properties of the mediator, such as its mass (provided it is kinematically allowed) and spin (provided it is allowed by spin-statistics). This is because the daughters inherit the boost of the mediator. The mediator boost is $m_\chi/m_Y$, and daughter particles have energies that are fixed fractions of $m_Y$, so $m_Y$ cancels. However, this can vary for gamma rays and neutrinos made through pions.  In the latter case, there are variable numbers of pions, with different fractions of energy going to gamma rays, etc., so there can be some variance. This is observed in particular for the different mediator mass and the $4b$ final state, owing to more hadronic cascade decays being available with higher mediator mass 
and consequently softening the spectra (this behavior for gamma-ray spectra is consistent with Ref.~\cite{Elor:2015tva}). For direct decays to gamma rays and neutrinos, the low-energy bound of the box spectra depends on the mediator mass~\cite{Ibarra:2012dw}, but this is only significant if the mediator is not sufficiently boosted. Also note that for mediator decay to gamma rays, some lower energy gamma rays can be produced from radiated electrons. However, these small differences do not provide any appreciable differences to our results, which predominantly arise from the high energy part of the spectrum.

\subsection{Optimal Signal Conditions}
\label{subsec:optimalsig}
\begin{figure}[h]
\centering
\includegraphics[width=\columnwidth]{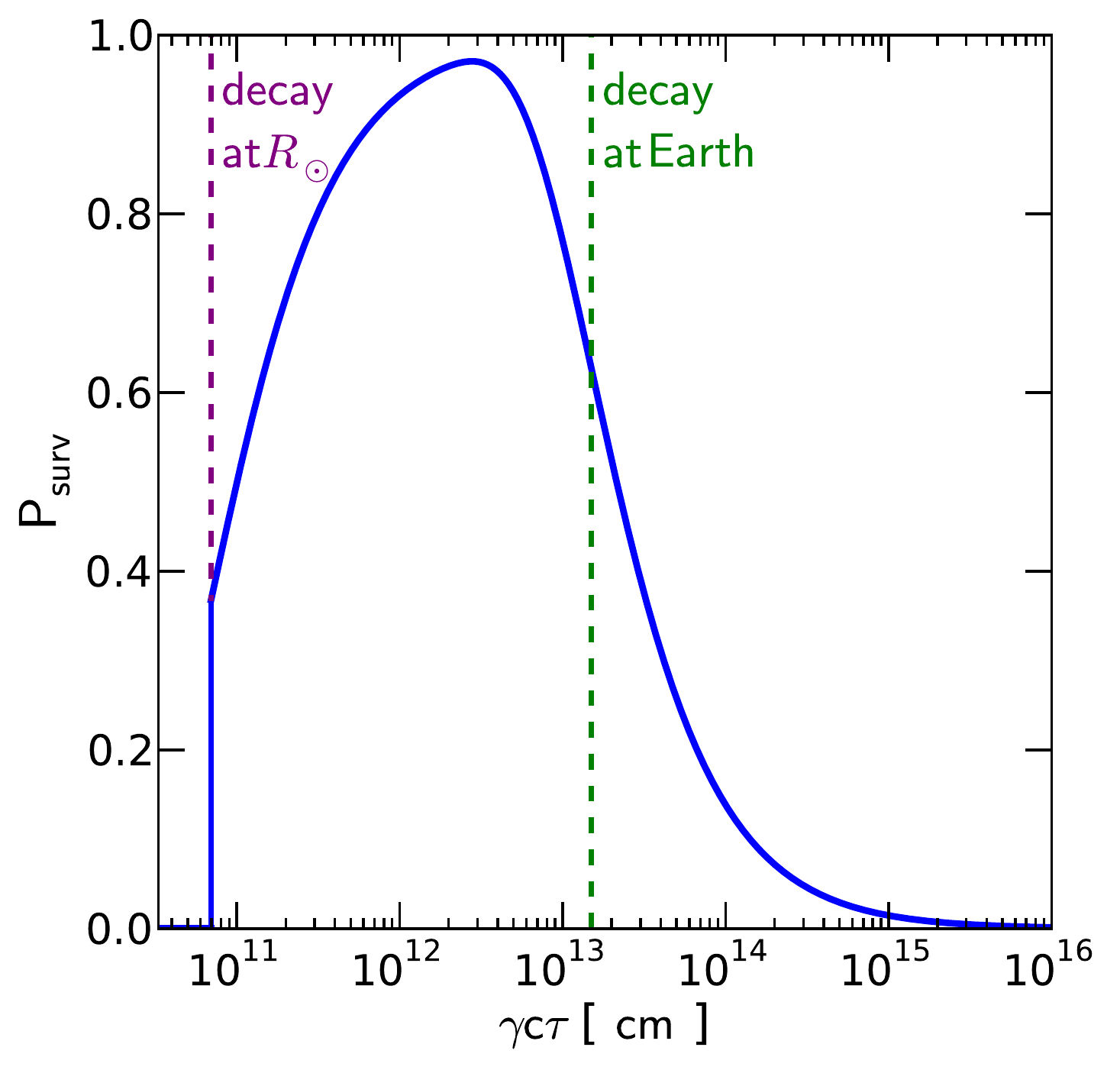}
\caption{Probability of gamma rays from the mediator surviving and reaching a detector at Earth, for varying mediator properties. This only takes into account decay exponentials, and assumes zero signal if $\gamma c \tau < R_\odot$. In the standard scenario, $\gamma c \tau \sim 0$, and the probability is exponentially suppressed for neutrinos due to parent-particle and neutrino absorption in the solar medium.}
\label{fig:pobs}
\vspace{-1mm}
\end{figure}

For decay products such as gamma rays to escape the Sun, it is required that the mediator $Y$ has a sufficiently long lifetime $\tau$ or sufficently large boost factor $\gamma=m_\chi/m_Y$, leading to a decay length $L$ that exceeds the radius of the Sun,
$R_\odot$, as
\begin{equation}
 L=\gamma\beta\tau  \simeq  \gamma c\tau > R_\odot,
 \label{eq:length}
\end{equation}
where $\beta$ is the speed of the mediator and $c$ is the speed of light. While the lifetime $\tau$ is related to the mediator mass $m_Y$, we just ensure combinations of the parameters are allowed by current constraints.

The probability of the signal surviving to reach the detector, $P_{\rm surv}$, provided the decay products escape the Sun, is
\begin{equation}
 P_{\rm surv}=e^{-{R_\odot}/{\gamma c\tau}}-e^{-{D_\oplus}/{\gamma c\tau}}.
\label{eq:pobs}
\end{equation}

Figure~\ref{fig:pobs} illustrates the survival probability for varying $\gamma c\tau$. In this work, we take $\gamma c \tau = R_{\odot}$.  The probability is relatively insensitive to $\gamma c \tau$, as survival probability is changing only by a factor \mbox{$\sim 2$}.  
For gamma rays, signal production is only possible if the mediator decays outside the Sun. 
For neutrinos, however, mediator decay inside the Sun provides a non-zero flux, but the signal is attenuated due to parent-particle and neutrino absorption. 
We assume mediators pass through Sun without attenuation, though such a feature is model-dependent.

We assume the signal strength only depends on $\gamma c \tau$. 
However, special scenarios can arise in some limiting cases.  
When $\gamma \gg 1$, the decay products are boosted and maintain a small opening angle. We focus on this case, where the Sun will appear to be a point source.  
When $\gamma c \tau \simeq R_{\odot}$ and $\gamma \sim 1$, mediators decay just outside the Sun and the Sun remains effectively a point source. 
However, when $R_{\odot} \ll \gamma c \tau < D_{\oplus}$ and $\gamma \sim 1$, the decay products would appear to be a halo around the Sun.  Typically diffuse-emission sensitivity is worse than that of point sources, and the analysis is more involved~\cite{2011ApJ734116A, Ng:2015gya, Zhou:2016ljf}. Thus we do not consider this case. Lastly, the Sun can absorb some of the gamma rays produced by the mediators.  This only occurs when the decay length is small and the mediators have a boost component away from the observer. For typical mediator masses and boost factors, only the low-energy part of the spectrum is affected; hence our results are not affected.

Therefore, our premise assumes a high mediator boost that requires the mediator to be sufficiently lighter than the DM mass. This is easily obtained across a range of DM masses for direct decays to gamma rays, neutrinos and electrons. 
For heavier final states such as taus and $b$-quarks, larger DM masses would be required to produce a highly boosted mediator that could kinematically produce such final states. 
As there is not a hard cutoff for such criteria, we show sensitivity of gamma rays and neutrinos for all DM masses that could produce such final states, even if the mediator would not be highly boosted, but potential weakening of sensitivity due to such directional loss in such regions should be kept in mind. 

Lastly, we neglect the extra gamma-ray component from secondary electrons inverse-Compton scattering with the ambient photons~\cite{Moskalenko:2006ta, Orlando:2006zs}.  This component is heavily suppressed due to the anisotropic solar photon distribution~\cite{Moskalenko:1998gw, Orlando:2013pza}. We also note that the gamma-ray contribution from DM annihilation in the solar WIMP (Weakly Interacting Massive Particle) halo outside the Sun is negligible~\cite{2010PhRvD81f3502S}.

\section{High-Energy Solar Gamma Rays}
\label{sec:limits_gamma}

In this section we discuss our procedure and results for long-lived dark mediators using solar gamma rays with Fermi-LAT, HAWC, and LHAASO.
 
\subsection{Procedure}

Fermi-LAT analyses provide the best measurements of solar gamma rays. In 2011, Fermi detected \mbox{$0.1$--$10$~GeV} solar gamma rays, measuring an energy flux \mbox{$\sim10^{-8}$ GeV cm$^{-2}$ s$^{-1}$ \cite{2011ApJ734116A}}. Since then, Fermi has collected more data and improved the data quality. This updated Fermi data are analyzed in Ref.~\cite{Ng:2015gya}, where the results are extended to 100 GeV solar gamma rays, measuring energy fluxes of $\sim 10^{-8}$ GeV cm$^{-2}$ s$^{-1}$. Together, these analyses provide much improved observational studies of solar gamma rays, which have not been fully explored in the context of long-lived mediators. For higher energy gamma rays ($\sim10^2-10^5$ GeV), HAWC and LHAASO could be used to observe the Sun~\cite{Ng:2015gya, Zhou:2016ljf}, but this has not yet been exploited. 

In this work we demonstrate that current Fermi-LAT analyses can be used to set strong limits through solar gamma rays from long-lived mediators. We also demonstrate that upcoming analyses from HAWC~\cite{Abeysekara:2013tza,Abeysekara:2017mjj} and LHAASO~\cite{Zhen:2014zpa} are extremely sensitive to solar gamma rays from long-lived mediators.

\begin{figure*}
     \begin{center}
        \subfigure{
            \includegraphics[width=\columnwidth]{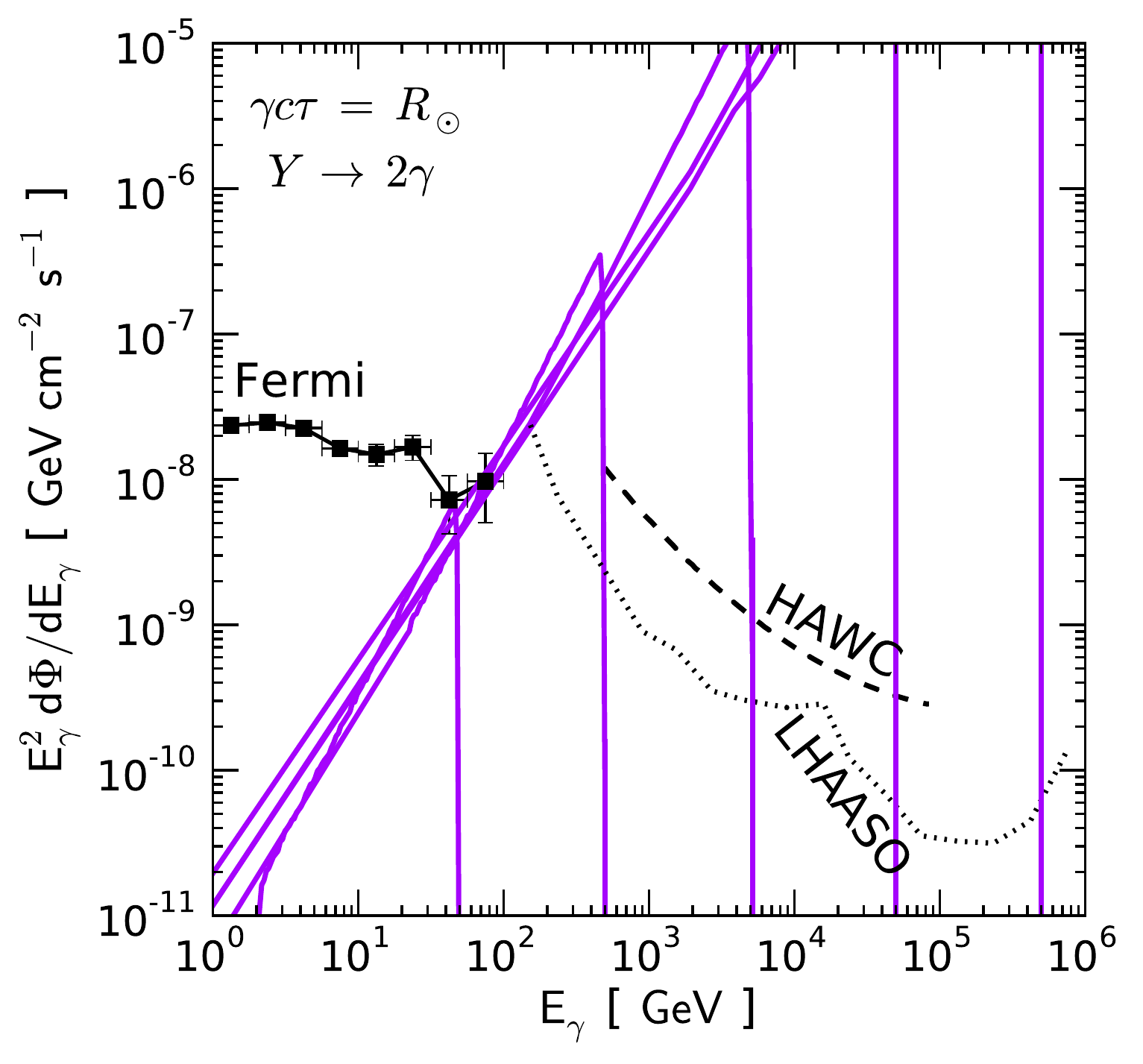}}
        \subfigure{
           \includegraphics[width=\columnwidth]{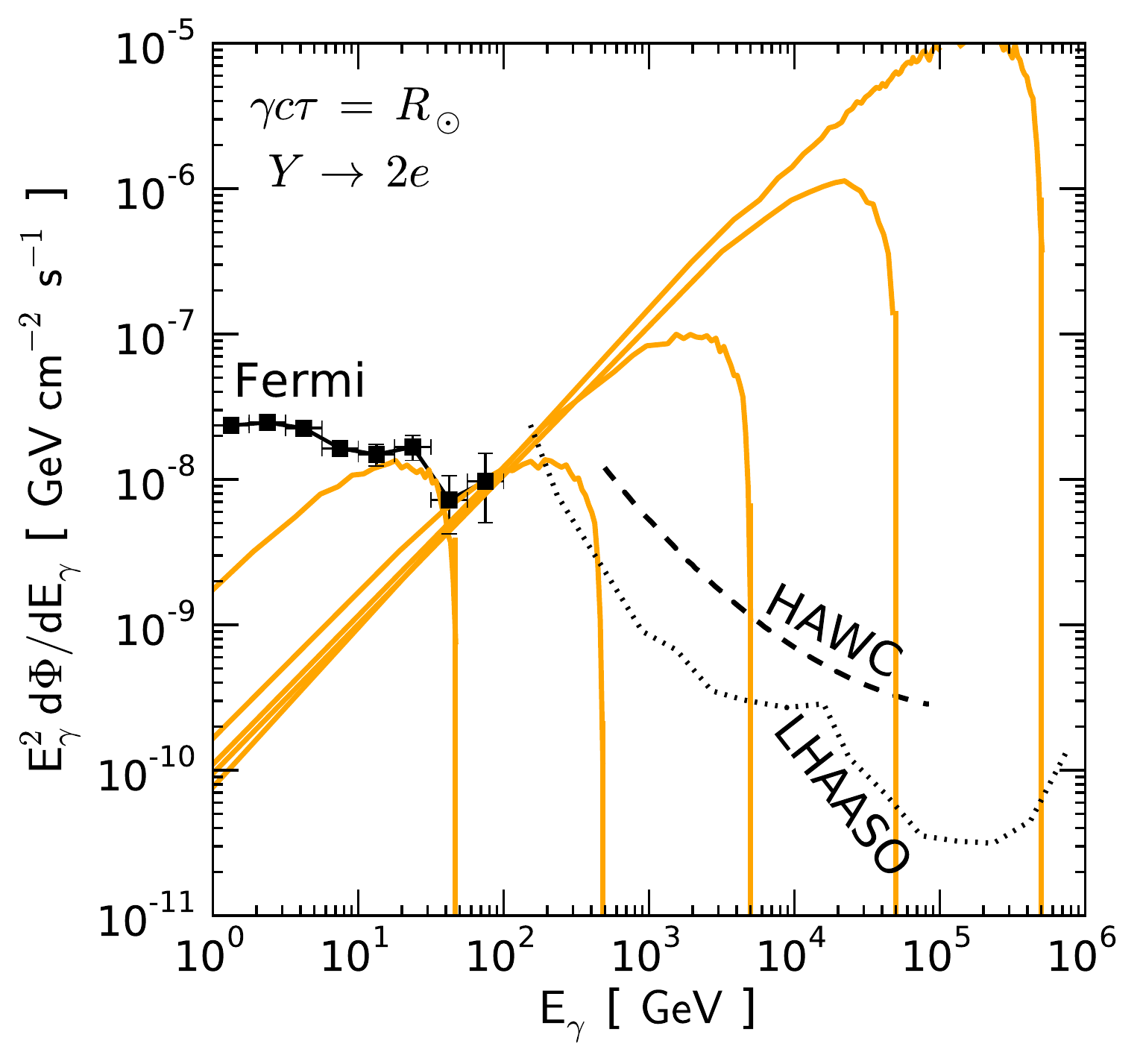}}
        \\ 
        \subfigure{
            \includegraphics[width=\columnwidth]{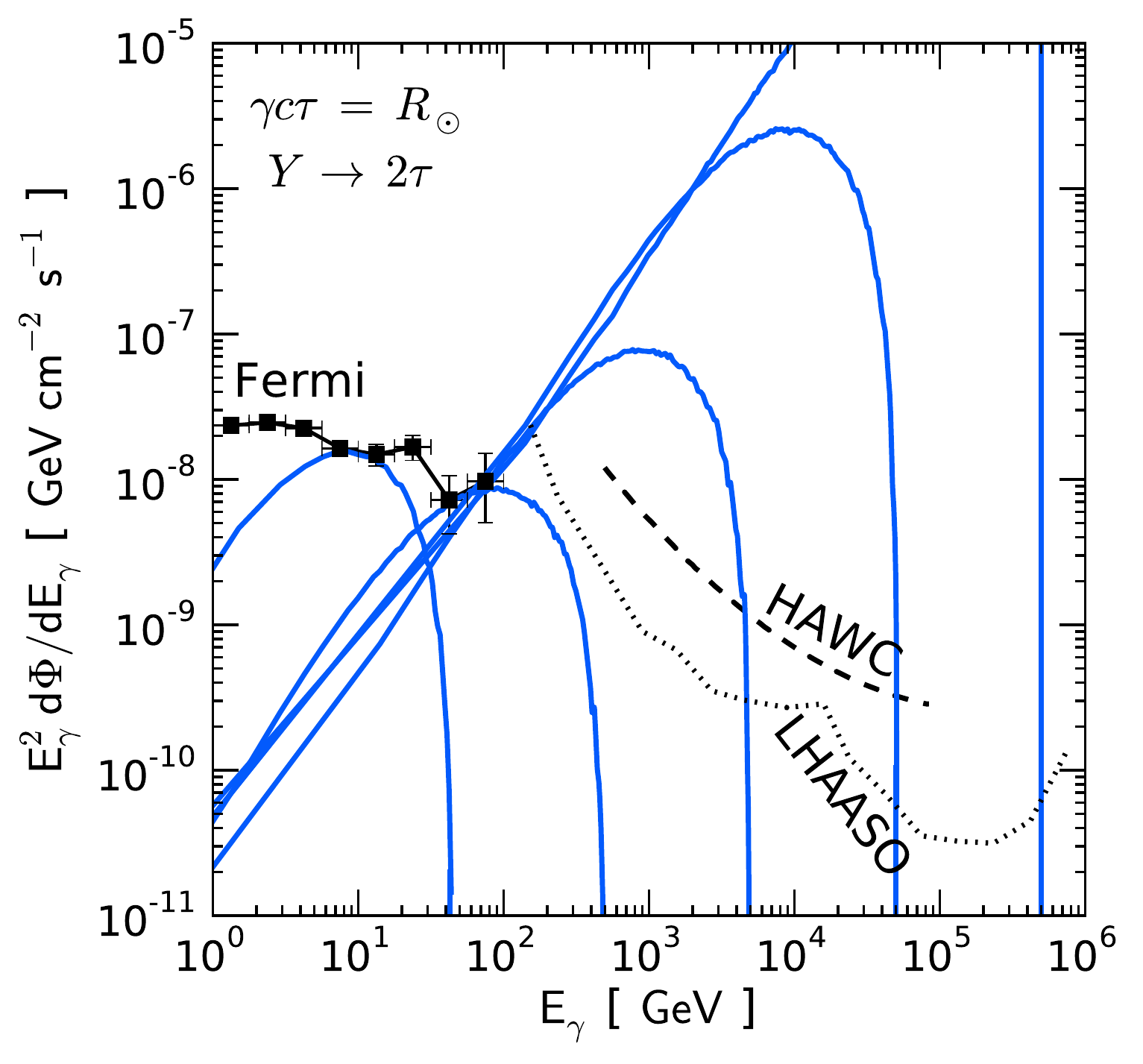}}
        \subfigure{
            \includegraphics[width=\columnwidth]{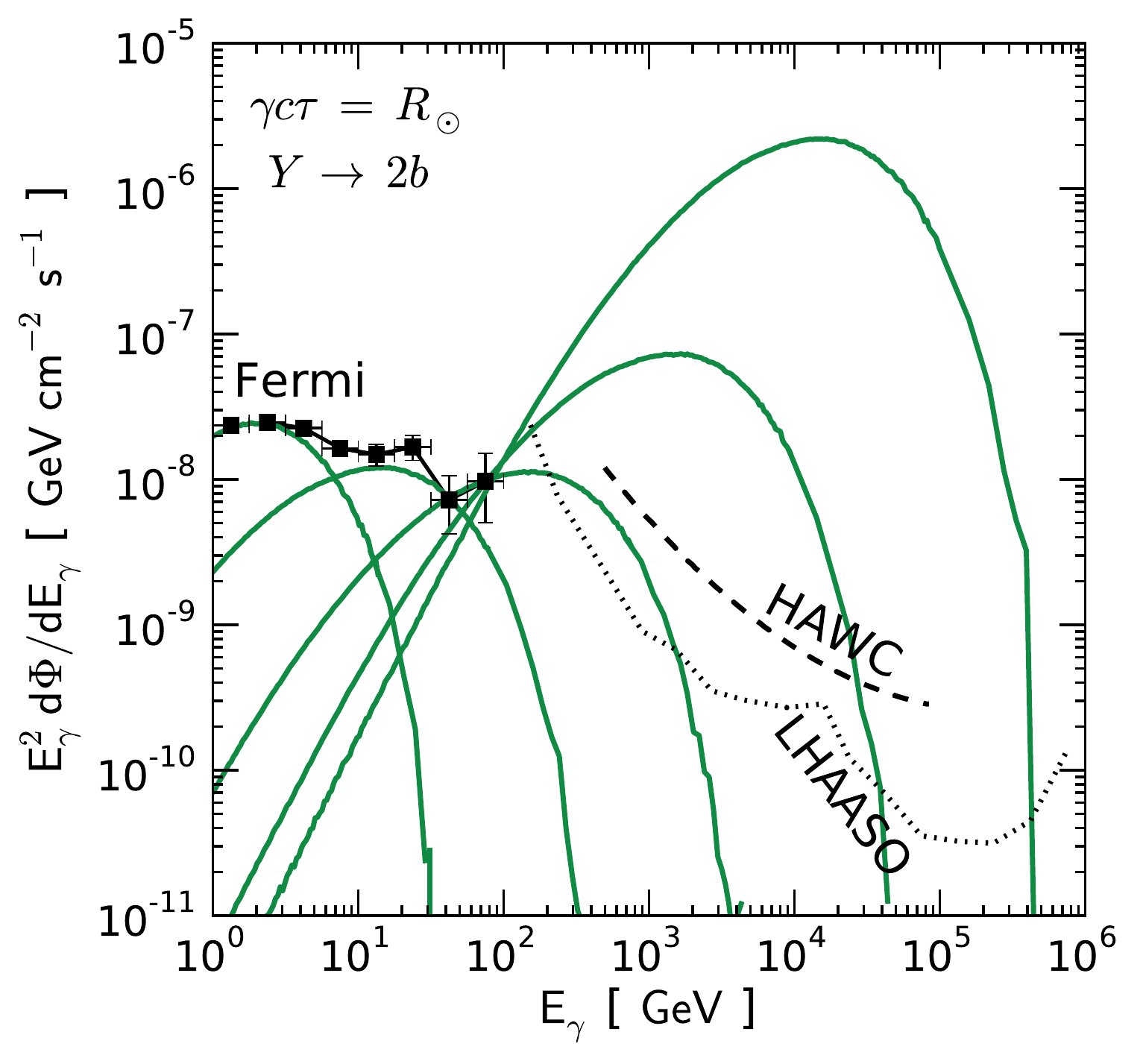}}
    \end{center}
    \caption{Estimates of presently allowed gamma-ray spectra from solar observations by Fermi-LAT, for final states as labeled and for DM masses of $m_\chi=50,\, 500,\, 5000,\,50000,\,500000$ GeV (left to right), with $\gamma c\tau=R_\odot$. The fluxes shown have been scaled by different annihilation rates for each mass and final state, such that they reach the sensitivity limit. HAWC and LHAASO do not yet provide constraints, but can do so soon.}
   \label{fig:flux_examples}
\end{figure*}

Figure~\ref{fig:flux_examples} illustrates how our new limits are obtained from existing Fermi-LAT data. For a fixed branching fraction and $P_{\rm surv}$, the spectra $E_\gamma^2 dN/dE_\gamma$ generated are scaled with arbitrary increasing annihilation rate $\Gamma_{\rm ann}$. Once the energy flux exceeds the sensitivity of \mbox{Fermi-LAT} in any energy bin, an upper limit on the value of $\Gamma_{\rm ann}$ from Fermi-LAT is obtained. Future HAWC and LHAASO analyses will also have strong sensitivity to this scenario, as we are the first to show.

\begin{figure*}
     \begin{center}
        \subfigure{
            \includegraphics[width=\columnwidth]{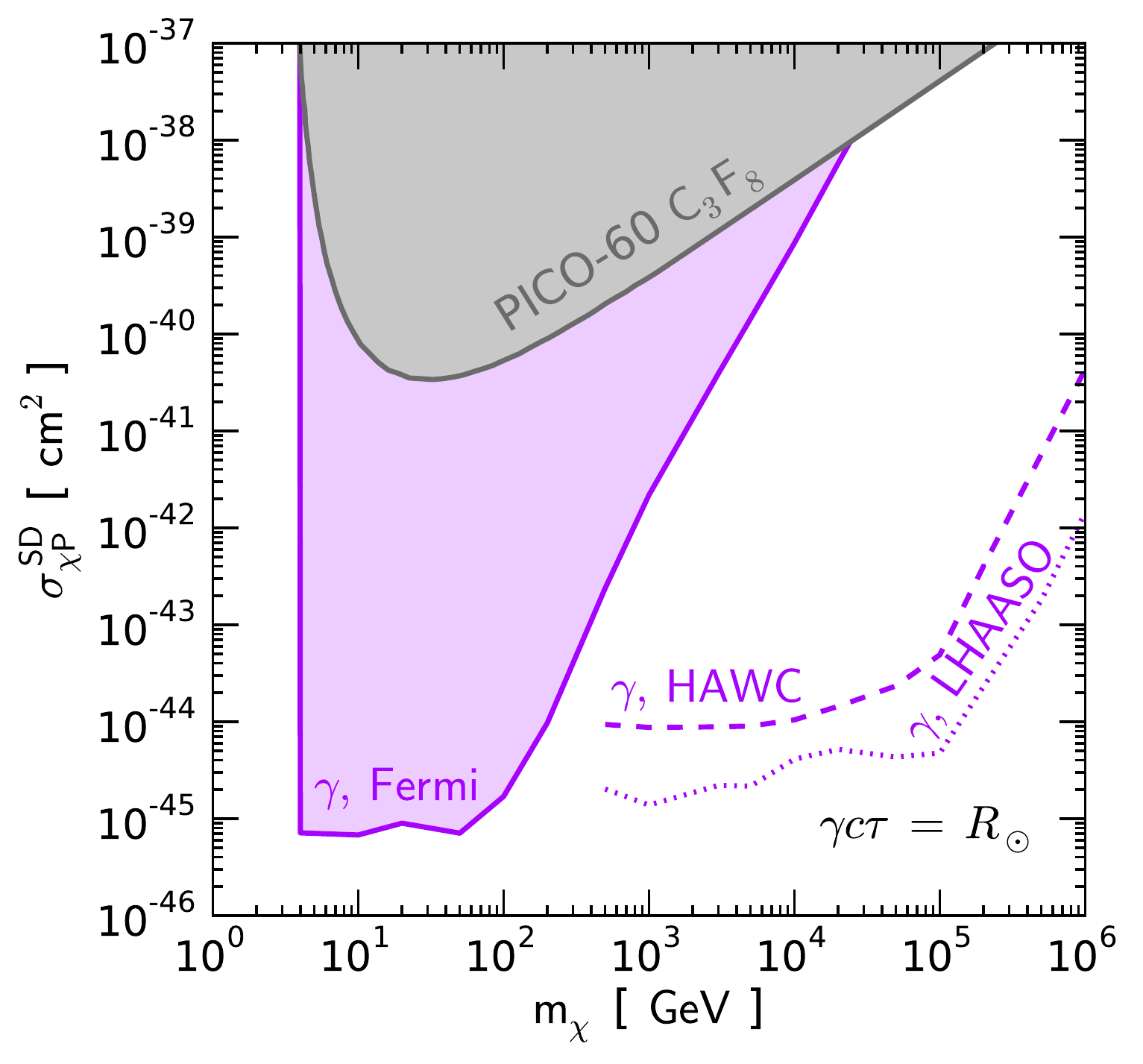}}
        \subfigure{
           \includegraphics[width=\columnwidth]{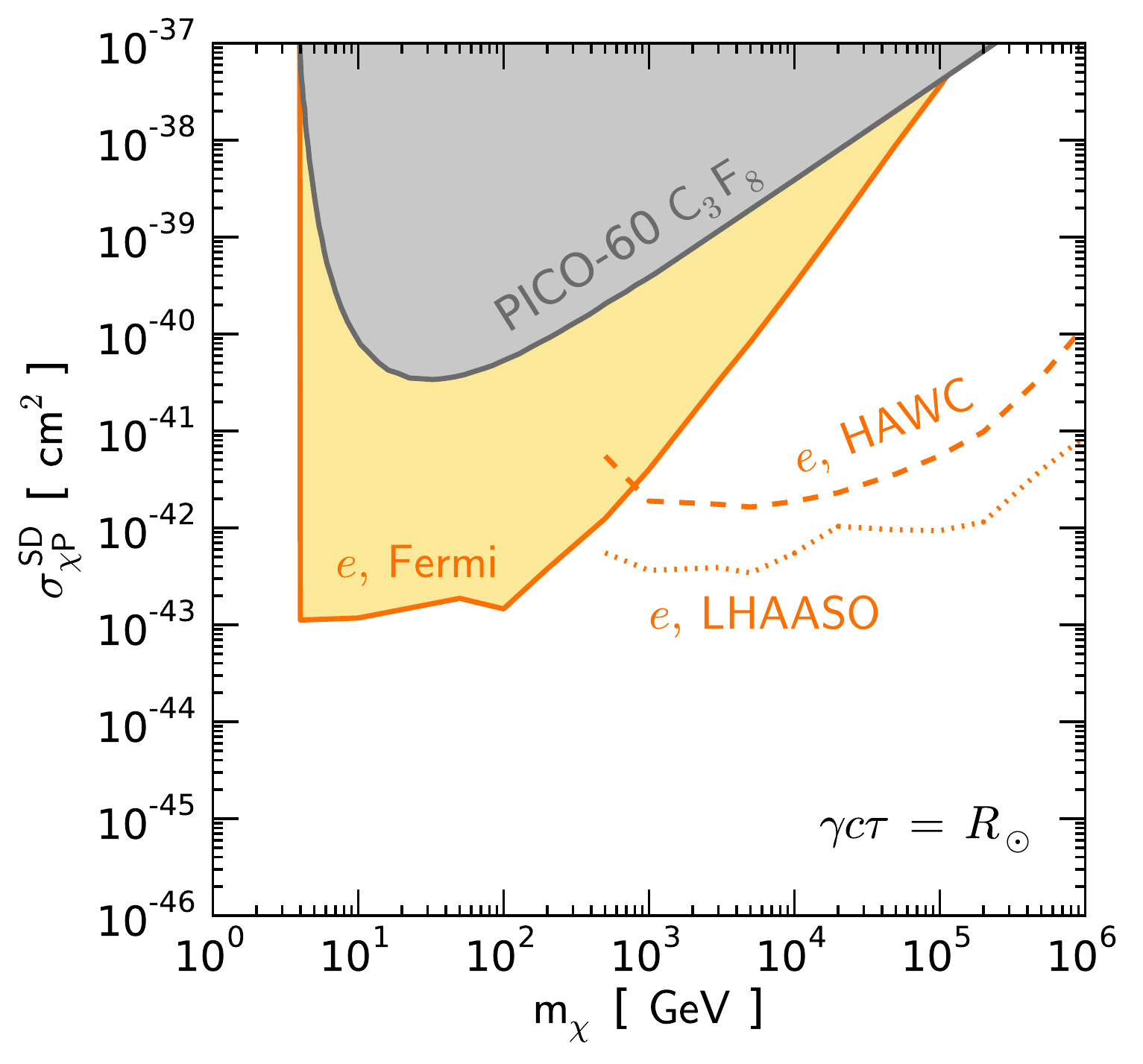}}
        \\ 
        \subfigure{
            \includegraphics[width=\columnwidth]{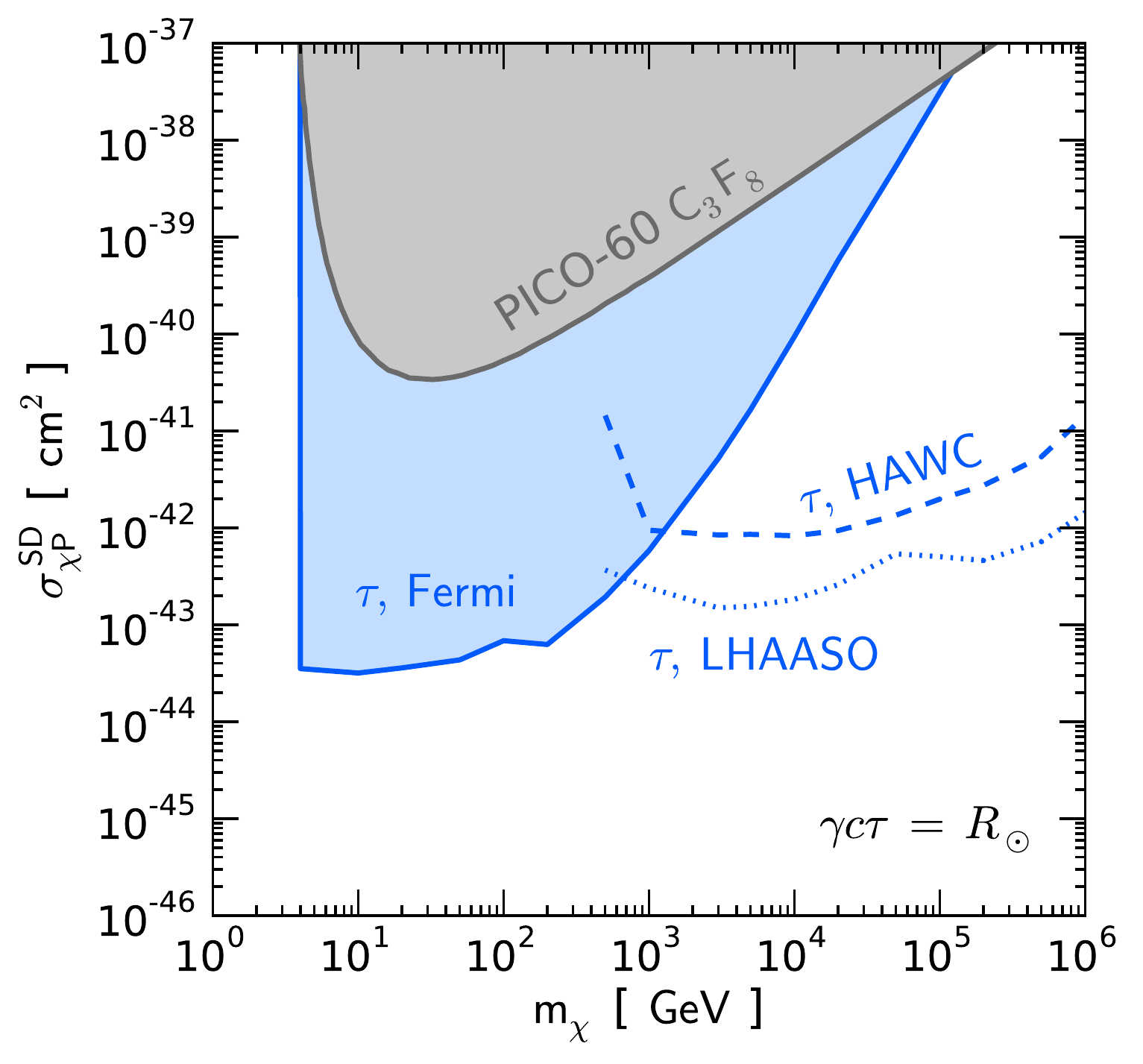}}
        \subfigure{
            \includegraphics[width=\columnwidth]{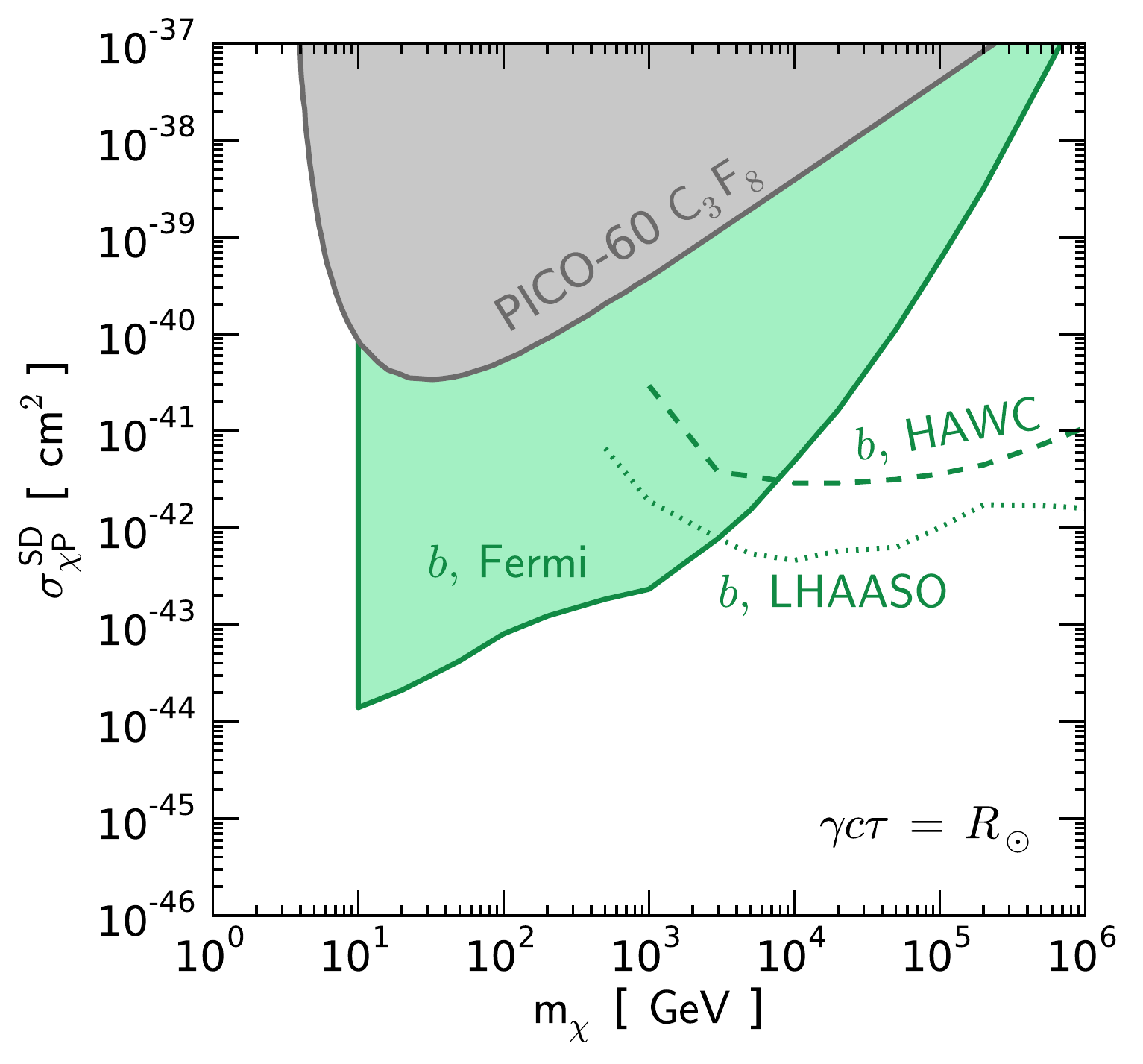}}
    \end{center}
    \vspace{-1mm}
    \caption{Optimal sensitivity for DM scattering cross sections from current and future solar gamma-ray observations, for DM in the Sun annihilating to pairs of long-lived mediators decaying to the particles labeled. Here the mediator decays just outside the Sun ($\gamma c\tau=R_\odot$). Our new limits from Fermi-LAT solar gamma-ray data are shown (shaded, solid), while our calculations of the estimated 1-year sensitivity from HAWC (dashed) and LHAASO (dotted) can be tested in future analyses.  PICO-60 C$_3$F$_8$ \cite{Amole:2017dex} $90\%$ C.L. limits are shown in gray. See text for details about the model assumptions
for the limits and sensitivities.} 
   \label{fig:SD_limit_photons}
\end{figure*}

Figure~\ref{fig:SD_limit_photons} illustrates our new limits (for Fermi-LAT) and our calculated sensitivities (for HAWC and LHAASO) to the DM scattering cross section using solar gamma rays, for mediator decay just outside the Sun ($L=\gamma c\tau=R_\odot$, implying $P_{\rm surv}\approx0.4$).
An upper limit on the annihilation rate implies an upper limit on the scattering cross section, which we obtain using {\sc DarkSUSY}~\cite{Gondolo:2004sc} as described in the previous section. As it is difficult to be competitive with strong direct detection limits on the DM spin-independent scattering cross section, we only show the spin-dependent results.

\vspace{-1mm}

\subsection{Discussion of Results}

At high mass, the Fermi sensitivity weakens due to the scaling of the capture rate ($\propto m_\chi^{-2}$) and due to the peak of the spectrum moving out of its energy range. This is why the sensitivity limits for final states with harder spectra, such as direct decay to gamma rays, weaken faster than softer spectra, such as those from $b$-quarks. For the softer spectra, this also leads to Fermi being more sensitive than HAWC and LHAASO for some higher DM masses, even in the 1--10 TeV DM mass range.  With HAWC and LHAASO, there is good sensitivity at high DM mass due to the increased energy range and flux sensitivity relative to Fermi.

The optimal long-lived mediator sensitivities with gamma rays shown in Fig.~\ref{fig:SD_limit_photons} are extremely powerful, outperforming the best spin-dependent direct detection limits from PICO by several orders of magnitude. Low DM masses are particularly promising with Fermi --- in the optimal scenario the sensitivity in the $m_\chi\approx 20-100$ GeV region outperforms the best spin-dependent direct detection experiments by about six orders of magnitude. HAWC and LHAASO are similarly powerful at high DM masses.

Furthermore, for some final states the optimal sensitivity with solar gamma rays even outperforms the projected sensitivity of upcoming direct detection experiment DARWIN, which is predicted to be sensitive down to spin-dependent scattering cross sections of $\sigma^{SD}_{\chi P}\approx 10^{-43}$ cm$^{-2}$~\cite{Aalbers:2016jon}. This means that the best probe of the DM spin-dependent scattering cross section in the near future may be from solar gamma rays, if the dark sector contains a long-lived dark mediator.

Again, it is important to note that these are the optimal sensitivities: we have assumed decay immediately outside the Sun and a $100\%$ branching fraction to the particles detailed in the plots, and kept to an accuracy of a factor of a few. 
However, while there will certainly be factors that degrade the sensitivity, but they will certainly be less than the gain from allowing mediators to escape the Sun, which for gamma rays allows a non-zero flux, and for neutrinos lifts the exponential suppression of the flux due to attenuation in the Sun. Discussion of interpretation of these results in the context of some models is in Sec~\ref{sec:models}.

\section{High-Energy Solar Neutrinos}
\label{sec:limits_neutrino}

In this section we discuss our method and results for long-lived dark mediators using high-energy neutrino observations with neutrino telescopes. 

DM annihilations in the Sun produce neutrinos that could be detectable.  In this case, muon neutrinos are the most relevant, as the final state muons retain much of the directionality, which is essential for suppressing the atmospheric neutrino background. 
Searches for high-energy neutrinos from the Sun with neutrino telescopes provide the strongest limits on the DM-proton spin-dependent scattering cross section~\cite{Aartsen:2012kia,Choi:2015ara, Aartsen:2016exj}.  A search has also been conducted by Antares for some specific long-lived mediator channels~\cite{Adrian-Martinez:2016gti}. 

In models with long-lived mediators, the sensitivity is enhanced~\cite{Bell:2011sn} compared to the case where neutrinos are promptly produced at the core of the Sun. 
These enhancements can be understood by two considerations: \\

\noindent {\bf (1) Less cooling of the secondaries.}  For dark matter annihilation in the Sun with short-lived mediators, high-energy neutrinos are produced inefficiently, as they come only from the rare particles, such as gauge bosons or heavy mesons, that decay before losing energy.  (The more common particles, for example pions and kaons, lose energy and decay at rest, producing only low-energy neutrinos~\cite{Rott:2012qb,Bernal:2012qh}.)  However, if the mediators escape the Sun, the neutrinos from pions and kaons will be emitted at high energy, substantially increasing the flux.\\

\noindent {\bf (2) Less neutrino absorption from the solar matter. } High-energy neutrinos~($> 100$\,GeV) produced at the core of the Sun are exponentially suppressed due to absorption from the solar matter.   If the mediators decay outside the core, beyond which the density falls exponentially, this suppression is lifted and the high-energy neutrino flux is greatly enhanced.  This is especially important as neutrinos with higher energies are more detectable, due to increased cross section and decreased backgrounds.\\

These enhancements are especially significant for high-mass DM, where the secondary multiplicity is large and neutrino absorption is important.  Therefore, we focus our discussion on large neutrino telescopes such as IceCube and KM3NeT.  In any case, except for pure neutrino final states, the sensitivity to gamma rays~(Sec.~\ref{sec:limits_gamma}) is much stronger than that for neutrino detectors such as Super-K.

\subsection{Procedure}

We first consider the neutrino flux from DM annihilations through long-lived mediators. 
The muon neutrino flux at Earth is obtained from $(\nu_{e}, \nu_{\mu},\nu_{\tau})$ at production~(\ref{sec:spectra}) multiplied with the weighting (0.27, 0.35, 0.38) due to mixing.  For pure neutrino channel, we assume equal flavor ratio at production. 
The weighting assumes the neutrinos arrive as an incoherent mixture of mass eigenstates~(mixing angles are obtained from Ref.~\cite{Olive:2016xmw}) due to vacuum mixing.  
We ignore the matter effect as we focus on mediators that decay outside the Sun.  
We also note that the oscillation length can approach 1\,AU at $\sim10$\,TeV and $\sim1$\,PeV, which we ignore as we will integrate the spectrum over large energy bins. 
These are good approximations for most of the energy range that we consider, especially given that we aim for an accuracy of a factor of a few. 

\begin{figure}[t]
\centering
\includegraphics[width=\columnwidth]{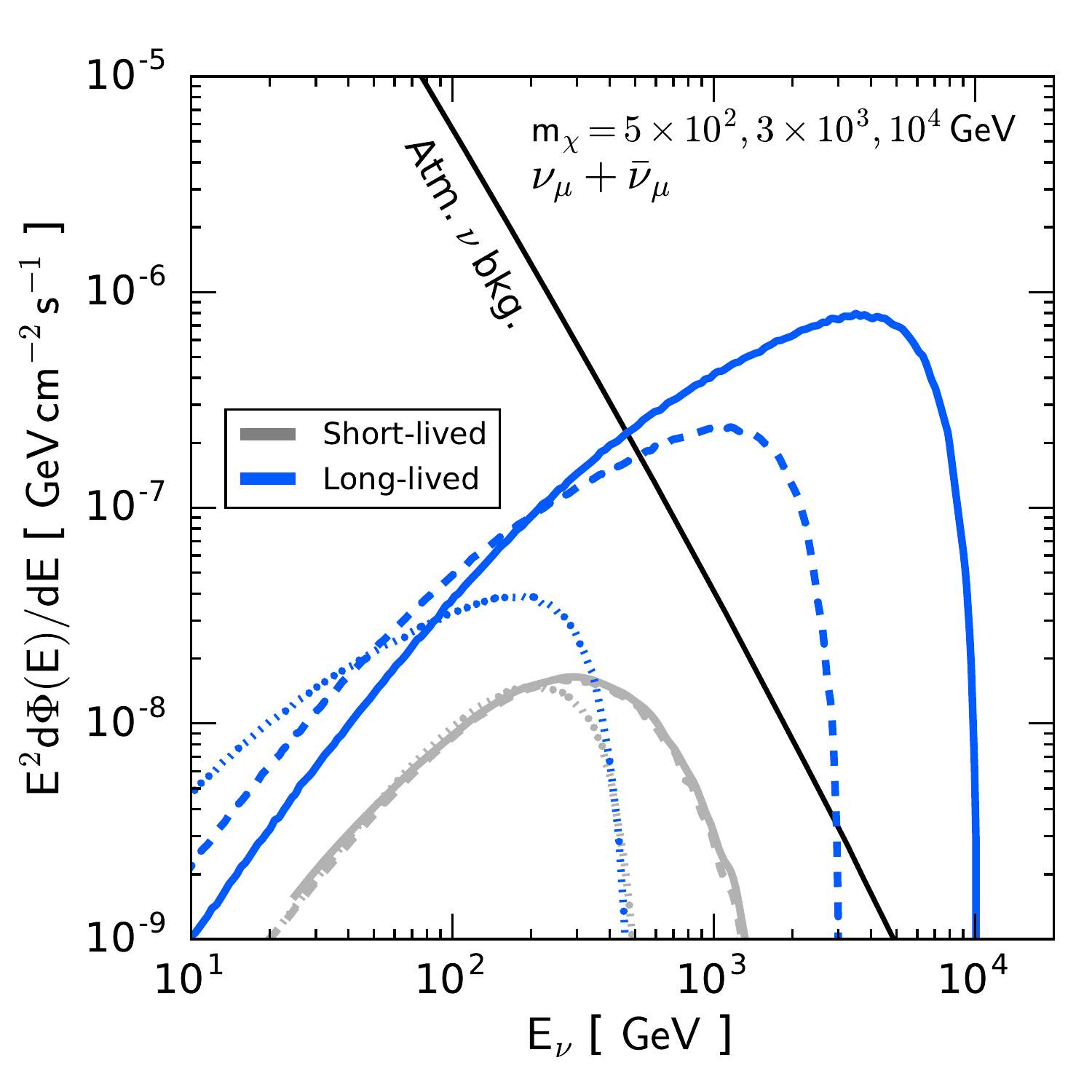}
\caption{ Neutrino flux from DM annihilating in the Sun to long-lived mediators with $Y\rightarrow 2\tau$ and $P_{\rm surv}$ is set to 1 for easy comparison~(only in this figure). Also shown are the cases for short-lived mediators in the center of the Sun with $\chi\chi\rightarrow\tau\bar{\tau}$ and the atmospheric background within the neutrino--muon opening angle. The dotted, dashed, and solid lines correspond to DM masses $5\times 10^{2}, 3\times 10^{3}$, and $10^{4}$\,GeV, respectively.   The annihilation rate is $10^{18}\,{\rm s^{-1}}$ for all DM masses. The neutrino flux for long-lived cases is enhanced, especially for large $\rm m_{\chi}$. }
\label{fig:neutrino_flux}
\end{figure}

Figure~\ref{fig:neutrino_flux} shows the muon neutrino fluxes where DM annihilates through long-lived mediators, \mbox{$Y\rightarrow2\tau$}, with $P_{\rm surv} = 1$~(only in this figure) for easy comparison. 
We compare them to the muon neutrino flux from \mbox{\sc WimpSim}~\cite{Blennow:2007tw}, for $\chi\chi\rightarrow\tau\bar{\tau}$ at the center of the Sun~(noted as the ``short-lived'' case).  

With long-lived instead of short-lived mediators, the neutrino fluxes are larger due to less energy loss of neutrino-producing secondaries and less attenuation of the neutrinos.
The spectra are also slightly softer, due to decaying into 4 final states instead of 2. 
While the short-lived cases are all exponentially attenuated above about 100\,GeV, the long-lived cases have significantly higher flux at higher energies, which improves the sensitivity.

For comparison, in Fig.~\ref{fig:neutrino_flux}, we also show the atmospheric neutrino flux, which is the dominant background for solar DM searches.  We use the all-sky averaged intensity from Ref.~\cite{Honda:2015fha} from the South Pole, and use the parametric form in Ref.~\cite{Sinegovskaya:2014pia} to extrapolate to high energies, after matching the normalization.  The background flux is estimated by considering neutrinos within the $\nu - \mu$ opening angle, $\theta_{\nu\mu}\simeq 1^{\circ}\sqrt{E_{\nu}/1\,{\rm TeV}}$.

To estimate the sensitivity, we compute the muon spectrum from neutrino charge-current interactions using the described neutrino fluxes.  
The average muon energy, $\langle E_{\mu}\rangle$, is related the neutrino energy, $E_{\nu}$, by $\langle E_{\mu}\rangle = E_{\nu}(1- y )$, where $ y $ is the average inelastic parameter~\cite{Gandhi:1998ri, Gandhi:1995tf}. For simplicity, we assume $y = 0.4$ throughout our energy range of interest, and ignore the distribution of the final state muons and take $E_{\mu}  = \langle E_{\mu}\rangle$.  

The muons can be detected as entering muons, when the interactions occur outside the detector volume. Taking into account the energy loss and the simplified assumption above, the differential rate is~\cite{Gaisser:1990vg, Kistler:2006hp}
\begin{equation}
\frac{dN}{dE_{\mu}} \simeq \frac{N_{A} \rho A_{\rm eff}^{\mu} T}{ \rho \left(\alpha  + \beta E_{\mu}\right) } \int_{ \frac{E_{\mu}}{1-y} }^{\infty}  dE_{\nu} \frac{d\Phi}{dE_{\nu}}(E_{\nu}) \sigma(E_{\nu})\, ,
\end{equation}
where $d\Phi/dE_{\nu}$ is the neutrino flux, $\sigma$ is the interaction cross section~\cite{Gandhi:1998ri, Gandhi:1995tf},  $N_{A} = 6.02\times 10^{23}$ is the Avogadro number, $\rho \simeq 1\,{\rm g\,cm^{-3}}$ is the density, $A^{\mu}_{\rm eff}$ is the effective detecting area of muons, $T$ is the exposure time of the detector, $\alpha = 2.0 \times 10^{-6}\,{\rm TeV\, cm^{2}\, g^{-1}}$, and $\beta = 4.2 \times 10^{-6}\, {\rm cm^{2}\,g^{-1} }$.
 
The muons can also be detected as starting muons, when the interactions occur inside the detector volume.  The  differential rate is
 \begin{equation}
\frac{dN}{dE_{\mu}} \simeq {N_{A} \rho V T}\frac{1}{1-y} \left[\frac{d\Phi}{dE_{\nu}}(E_{\nu}) \sigma(E_{\nu}) \right]_{E_{\nu} = \frac{E_{\mu}}{(1-y)}}\,,
\end{equation}
where $V$ is the fiducial volume of the detector. 

\begin{figure}[t]
\centering
\includegraphics[width=\columnwidth]{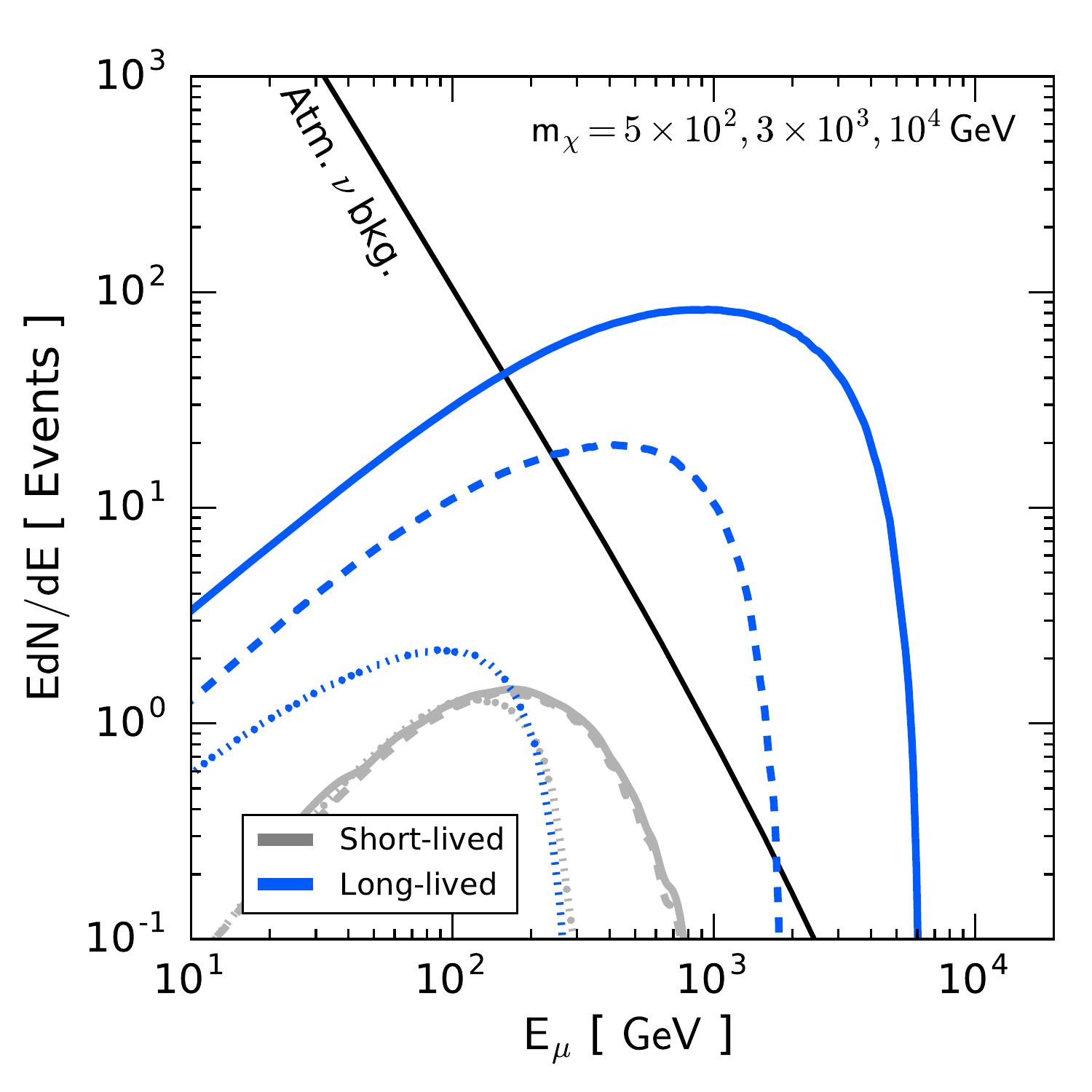}
\caption{ The muon spectrum~(entering + starting) for a gigaton neutrino detector with 317 days of exposure, obtained with the neutrinos fluxes from Fig.~\ref{fig:neutrino_flux}.  $E_{\mu}$ is defined as the energy of the muon when it first appears at the detector.   }
\label{fig:muon_flux}
\end{figure}

We consider an idealized gigaton scale detector, such as IceCube or KM3NeT, where $A^{\mu}_{\rm eff} = 1\, {\rm km^{2}}$ , $V= 1\,{\rm km^{3}}$.  We take the exposure to be $0.5\times317$ days, matching Refs.~\cite{Aartsen:2012kia, Aartsen:2016exj}.  The factor of $0.5$ comes from the fact that we only consider up-going muons, where the atmospheric muon background is greatly reduced. For upgoing events, neutrinos may be absorbed when they propagate through the Earth.  At the South Pole, the optical depth barely reaches unity when the Sun is at the lowest point below the horizon~(about $23^{\circ}$) for 1\,PeV neutrinos~\cite{Gandhi:1995tf}.  For our purpose and the mass range we consider, we can therefore safely ignore Earth absorption.  For a lower latitude detector, such as KM3NeT, this effect will be more important.

Figure~\ref{fig:muon_flux} shows the muon spectra that can be detected, using the neutrino fluxes from Fig.~\ref{fig:neutrino_flux}.   We note that the muon spectra for the short-lived cases start to be suppressed above about 100\,GeV.  
This is important as neutrino telescopes typically do not have good muon energy resolutions below a TeV.  The muon spectra are broader than the neutrino spectra because of the importance of entering muons, which lose energy outside the detector.

Finally, to estimate the sensitivity, we compute the number of signal and background events in two energy bins.  This is motivated by the realization that neutrino telescopes can estimate the muon energy above $\sim1$ TeV, when the muon energy loss becomes radiative~\cite{Aartsen:2013vja}.  The sensitivity is determined when the signal counts reach the background counts in either energy bin, similar to our gamma-ray analysis.  Here we also take $P_{\rm surv}$ to be $\sim0.4$~($\gamma c \tau = R_{\odot}$).  
There is some freedom in choosing the precise values for the energy bins. We find that the choice of $[10^{1.8}, 10^{3}]$\,GeV and $[10^{3}, 10^{6}]$\,GeV allows us to reproduce the IceCube limit~\cite{Aartsen:2016exj} up to factors of a few for the short-lived cases.  Our approach is simplifying: it is conservative to require the signal to be as high as the background; but this is compensated by the fact that we ignore the backgrounds from atmospheric muons, various detector effects, and reduction of signal efficiency from various data reductions~\cite{Danninger:2013sla}.  However, for our purpose of the estimating the improved sensitivity from long-lived mediators relative to the ``short-lived'' case, this is sufficient.

\begin{figure}
\centering
\includegraphics[width=\columnwidth]{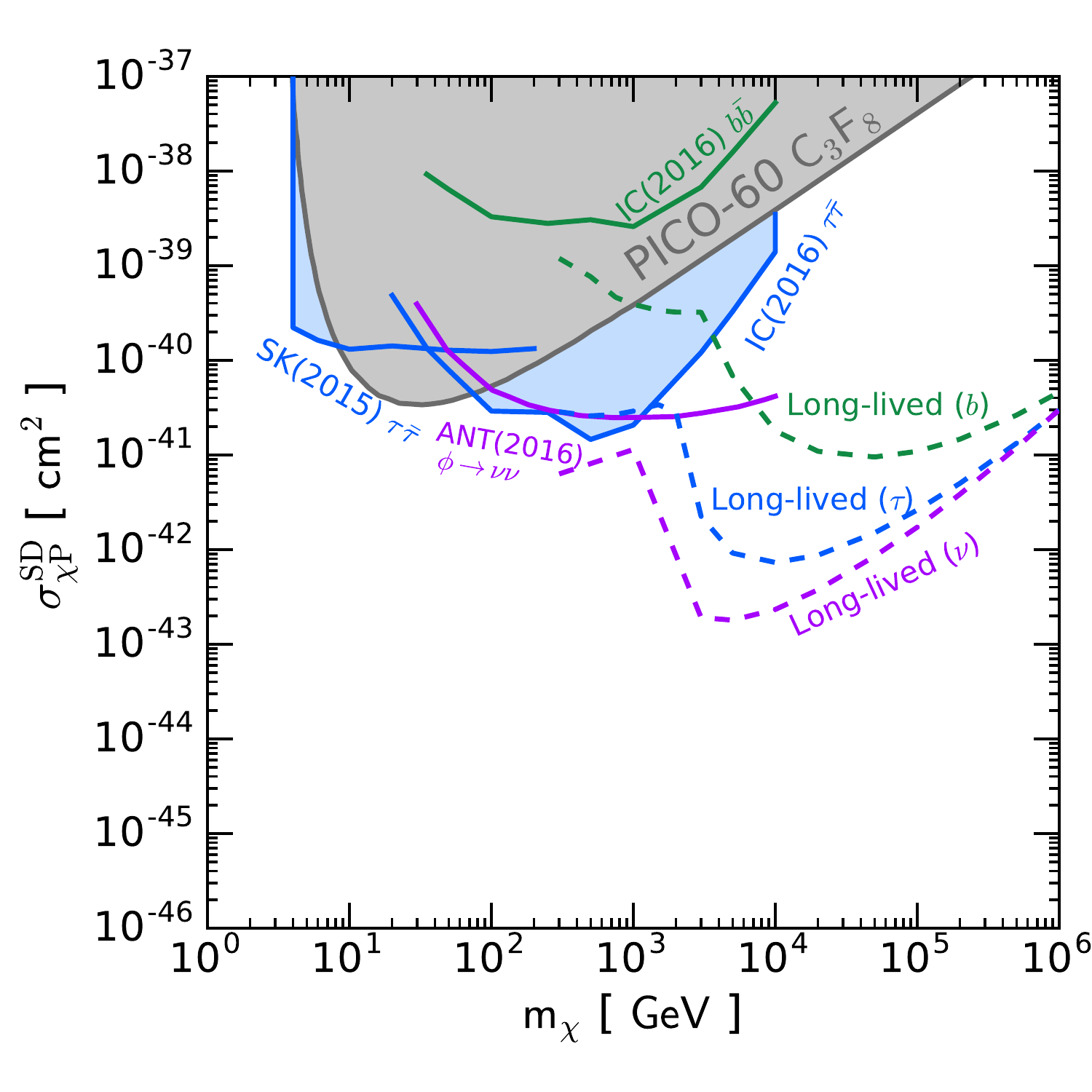}
\caption{Constraints and sensitivities for the spin-dependent DM scattering cross section.  
The dashed lines are the sensitivities for DM in the Sun annihilating to pairs of long-lived mediators that decay to the particles labeled~($\gamma c\tau=R_\odot$).  We also show current limits on short-lived mediators~(solid lines with shaded region) from Super-K~(SK), IceCube~(IC), \mbox{PICO-60} C$_3$F$_8$, as well as the limit from the search for secluded DM by Antares~(ANT). This highlights the significantly improved sensitivity that could be achieved by long-lived mediators. See text for details about the model assumptions for the limits and sensitivities.}  
\label{fig:neutrino_limit}
\end{figure}

\subsection{Discussion of Results}

Figure~\ref{fig:neutrino_limit} shows our estimated sensitivity compared with current constraints for standard WIMPs (short-lived case) from Super-K~\cite{Choi:2015ara} and IceCube~\cite{Aartsen:2012kia, Aartsen:2016exj}.   We also show the result obtained by Antares~\cite{Adrian-Martinez:2016gti}, which searched for secluded DM via the process \mbox{$\chi\chi\rightarrow YY\rightarrow \nu\bar{\nu}\nu\bar{\nu}$}.  We find that IceCube and KM3NeT can offer a significant improvement in sensitivity for the case of long-lived mediators, especially for high DM masses. For the $\tau$ final state, at lower masses, the long-lived mediator sensitivity is comparable to and even slightly weaker than the current limit.  This is expected from softer spectra and the $P_{\rm surv}$ factor. Much of the improved sensitivity comes the high-energy bin $> 1$\,TeV, which causes the kink near 1\,TeV.  Nominal WIMPs are not expected to produce such high-energy signals due to severe neutrino absorption in the Sun.  Hence, a 
detection of a high-energy muon from the Sun could signal the existence of long-lived mediators in the dark sector. 

As neutrino telescopes improve, DM searches from the Sun will eventually run into a sensitivity floor, due to the background flux of neutrinos produced by cosmic-ray collisions with the Sun~\cite{Arguelles:2017eao, Ng:2017aur, Edsjo:2017kjk}.  (This newly noted indirect-detection ``neutrino floor" is different than the direct-detection ``neutrino floor"~\cite{Billard:2013qya, Ruppin:2014bra}; the latter is caused by elastic scattering of MeV neutrinos produced in various sources, such as fusion in the Sun.)  The indirect-detection neutrino floor is a hard floor, because of the large present uncertainties in predicting the flux of solar atmospheric neutrinos.  In Ref.~\cite{Ng:2017aur}, it is shown that it is important to separate neutrino signals above and below about 1 TeV, and that this can be done by whether the muons they produce have radiative losses or not.  It is also shown that $>1$\,TeV muons from solar atmospheric neutrinos can be detected soon.  How could these 
be recognized as a signal of DM with a long-lived mediator?  A key test will be the associated gamma-ray flux, which is much larger for DM scenarios (see Fig.~\ref{fig:SD_limit_photons}) than for solar atmospheric interactions~\cite{Zhou:2016ljf}.

As mentioned above, for low mass DM ($<100$\,GeV), long-lived mediators do not offer much improvement to the sensitivity. In this case, gamma-ray observations by Fermi offer significantly larger potential discovery space.

\section{Model Interpretation of Results}
\label{sec:models}

While the purpose of this paper is to highlight the power of solar gamma rays and neutrinos to probe the DM parameter space in a pure phenomenological sense, rather than to be a complete study of DM models, in this section we briefly discuss potential interpretations of these results in the context of popular models. We caution that the limits shown in Figs.~\ref{fig:SD_limit_photons} and \ref{fig:neutrino_limit} are the optimal scenario, and other constraints should also be taken into account in model building (see Sec.~\ref{sec:constraints}). A specific model realization that saturates the limits is beyond the scope of this work.

\subsection{Dark Vector or Axial-Vector}

Spin-1 mediators cannot decay directly to two photons, by spin-statistics. Instead, final state photons may be obtained in other ways, such as electroweak bremsstrahlung, or hadronic decays.
Resulting gamma ray spectra are softer than direct decays, and so the sensitivity to gamma rays in such a scenario would be closer to the $b$ or $\tau$ channels. Of course, this is not a feature for the direct decay of a spin-1 mediator to neutrinos.

The dark photon, a gauge boson of a new $U(1)$ which kinematically mixes with SM hypercharge, is a popular spin-1 mediator. The dark photon can induce a large spin-independent scattering cross section, as the dark photon inherits Lorentz structures from kinetic mixing with the SM hypercharge, and it is difficult to remove the spin-independent contribution without fine-tuning cancellation by some other contribution. Therefore, competition with direct detection is a particularly important consideration in this scenario. Regardless, long-lived dark photon searches can still be more powerful than direct detection experiments~\cite{Ardid:2017lry}. 

Furthermore, there are other promising spin-1 mediator scenarios which do not have significant spin-independent scattering signals. For example, the spin-independent scattering rate can be suppressed if the scattering is predominantly inelastic, with a small mass splitting in DM states~\cite{Batell:2009zp,Smolinsky:2017fvb}. Alternatively, direct detection limits can be suppressed in an axial-vector mediator scenario, or in an (almost) hidden sector setup with larger DM couplings and smaller SM couplings.

\subsection{Fundamental Dark Scalar or Pseudoscalar}

In some models, particular decay channels dominate due to properties of the mediator. For example, a dark Higgs which couples to the SM Higgs via a Higgs portal term will predominantly decay to the heaviest decay product that is kinematically available. This can motivate a choice of a near $100\%$ branching fraction. In order to be sufficiently long-lived to escape the Sun, typically it is required that the mediator has very small couplings to its decay products, or alternatively there need to be a few orders of magnitude difference between the masses of the DM and the mediator. For a Higgs portal model, these requirements are easily met with a light mediator which consequently can only decay into light final states such as electrons. Alternatively, in a two Higgs-doublet Higgs portal model, the dark Higgs couplings may be uncorrelated with SM Higgs couplings, for example when one of the doublets is leptophilic.

Another promising scenario is DM annihilating to axions, or any light pseudoscalar. For the axion or axion-like particle (ALP) to be both sufficiently long-lived to escape the Sun and to produce a sizable gamma-ray flux, light decay products such as gamma rays or electrons could be directly produced. For the combination of parameters where the gamma-ray flux dominates, a sharp spectral feature would be observed. Full details of such a model are comprehensively discussed in Ref.~\cite{Batell:2009zp}. Also note that for pseudoscalar mediators, DM annihilation to a three-body pseudoscalar final state may be the dominant $s$-wave process, providing a different spectral shape~\cite{Abdullah:2014lla, Rajaraman:2015xka}.

\subsection{Multi-Mediator Scenarios}
In this paper, we consider the case that the dark sector contains a DM candidate and one dark long-lived mediator. Of course, the dark sector could be complex, with combinations of weak and strong dynamics, or several dark sector particles could be involved in the decay. If there exists a cascade of dark sector decays, the overall spectral shape will be softened. However, if there are only a few decays, this softening will not greatly affect the results~\cite{Elor:2015bho}. This also means that the dark sector particles involved in the decay can have shorter lifetimes than required to escape the Sun, and it is only required that the sum of the lifetimes are sufficiently long to escape the Sun.

Furthermore, more than one mediator present in a model is well motivated in some scenarios. For example, in the case there is a dark spin-1 boson with axial couplings, unitarity is violated at high energies unless a scalar is also included in the setup \cite{Cline:2014dwa, Kahlhoefer:2015bea}. When both a vector and scalar are present, the DM annihilation and indirect detection signals can be different, with DM annihilating into a vector plus scalar final state potentially dominating when kinematically allowed \cite{Bell:2016fqf,Duerr:2016tmh,Bell:2016uhg}. If the masses of both mediators are of the same scale, the sensitivity limits are not drastically different and depending on final states could approximately map to our results of DM annihilation to only one type of mediator \cite{Bell:2016fqf,Bell:2016uhg}. Such a scenario also produces a compelling way to produce a large gamma-ray flux while evading other constraints; only one of the mediators need to satisfy conditions to escape the Sun to provide 
some non-zero DM flux. In fact, a particularly promising scenario arises when the scalar is long-lived and escapes the Sun, and the vector does not \cite{Batell:2009zp}.

Generally, more than one mediator present can lead to destructive interference in direct detection signals, and consequently a blind spot in the direct detection limit. Interestingly, this may be covered by the solar spin-dependent limit we present on the scattering cross section instead.

DM may also exist in bound states, such as WIMPonium~\cite{MarchRussell:2008tu,Shepherd:2009sa,Braaten:2013tza,Laha:2013gva,Wise:2014ola,Wise:2014jva,Petraki:2014uza,vonHarling:2014kha,Petraki:2015hla,Laha:2015yoa}, or a dark pion, which can produce a large gamma-ray flux \cite{Bai:2015nbs}. In such a case self-interactions of the DM would be relevant~\cite{Zentner:2009is}, leading to a potentially varied relation between scattering, annihilation, and self-interaction rates, with equilibrium reached at a different time.
\vspace{-2mm}
\subsection{General Considerations}

The results we present are the most optimal case. Indeed, in many models, these sensitivity limits will be different. For example, there may be several decay modes of the dark mediator, reducing the branching fraction. In general, the sensitivity should be scaled accordingly. However, for a model with non-negligible direct gamma-ray decays, the gamma-ray spectra can be so sharp that it is what sets the limit across most of the parameter space (e.g., true for monochromatic gamma-ray lines, and for $4\gamma$ box spectra \cite{Ibarra:2012dw,Bringmann:2012ez}), and so a non-zero branching fraction to mixed final states may not affect the sensitivity within our accuracy of factor of a few. 

When applying results to specific models, relations between parameters such as decay width, lifetime, masses and decay length will vary, and will need to satisfy the conditions in Sec.~\ref{subsec:optimalsig}.
In general, it is easier to fulfill requirements on parameter combinations for neutrinos. As neutrinos can propagate from inside the Sun~\cite{Press:1985ug,Silk:1985ax,Krauss:1985aaa,Griest:1986yu,Gould:1987ww,Gould:1987ir,Gould:1991hx}, shorter decay lifetimes are allowed for neutrino detection, leading to less stringent constraints on relationships between mediator properties. In the case where the scenario is sub-optimal, the sensitivity needs to be scaled~(e.g., when $P_{\rm surv}\ll 1$ in the highly boosted case) or dedicated data analyses are required~(e.g., when the Sun is no longer a point source).

Lastly, to escape increasingly strong spin-independent direct detection limits, model building efforts are often constructed such that the spin-independent DM direct detection signal is either suppressed or non-existent, and only the weaker spin-dependent direct detection constraint is relevant. In this paper we have made the important observation that, even in such a case, very strong limits may arise on the spin-dependent scattering cross section by utilizing solar gamma rays or neutrinos, in the scenario that the dark mediator is long-lived.

\section{Other constraints}
\label{sec:constraints}

There exist other constraints relevant for a long-lived mediator setup, but they are mostly highly model dependent. In this section, for completeness we outline other relevant constraints, which would need to be considered in a complete analysis. These are:\vspace{2mm}
\begin{itemize}

\item {\bf BBN:} The observed relic abundance of SM particles by Big Bang Nucleosynthesis implies any new mediator generally should have a lifetime $\tau\lesssim1$s \cite{Chen:2009ab}. Note that depending on the model details, this can be relaxed~\cite{Pospelov:2010hj}.
\item {\bf CMB:} DM annihilation to SM products in the early universe is constrained by the Cosmic Microwave Background \cite{Adams:1998nr, Chen:2003gz, Padmanabhan:2005es, Slatyer:2015jla}.
\item {\bf Supernovae:} Relevant constraints may be obtained for mediators lighter than a GeV, from mediator decay and supernova cooling \cite{Dent:2012mx, Dreiner:2013mua, Kazanas:2014mca, Rrapaj:2015wgs, Zhang:2014wra}.
\item {\bf Colliders:} If the dark sector is secluded, limits from colliders may be negligible. Otherwise, limits may be set by LHC experiments Belle \cite{TheBelle:2015mwa} for a dark Higgs and dark photon, BaBar \cite{Lees:2014xha}, ATLAS~\cite{ATLAS-CONF-2016-042,ATLAS-CONF-2016-103} and CMS~\cite{Khachatryan:2016sfv}.
\item {\bf Beam Dump/Fixed Target experiments:} Beam dump and fixed target experiments are most relevant when the mediator has mass lighter than a GeV. Limits on mediator properties can be set from E137 \cite{Bjorken:1988as,Bjorken:2009mm}, LSND \cite{Athanassopoulos:1997er,Batell:2009di,Essig:2010gu} and CHARM \cite{Bergsma:1985is,Gninenko:2012eq}.
\item {\bf Other indirect detection signals:} Fermi-LAT and DES measurements of dwarf spheroidal galaxies can be relevant particularly at low DM mass \cite{Ackermann:2015zua, Fermi-LAT:2016uux}, and large positron signals~\cite{Kim:2017qaw} can be constrained by AMS-02 \cite{PhysRevLett.117.091103}. Also note that Fermi-LAT observed the Sun, searching for long-lived mediators directly decaying to electrons in the DM mass range 70--2000~GeV, which are stronger than the gamma-ray limits~\cite{Ajello:2011dq}.  In such cases, helio- and geo-magnetic field effects must be taken into account, especially at lower DM masses. 
\item {\bf Thermalization and Unitarity:} Thermalization can be important for $>10$ TeV DM, and unitarity issues exist for DM mass $\mathcal{O}(100)$ TeV \cite{Griest:1989wd,Hedri:2014mua} for a standard WIMP, which is reached at the edge of the DM mass range we consider. Furthermore bound state effects can be relevant if DM mass becomes too large \cite{vonHarling:2014kha}. 
\end{itemize}

\section{Conclusions}
\label{sec:conclusions}

It has long been known that high-energy neutrinos can be used to probe DM scattering and annihilation in the Sun.  If annihilation proceeds via long-lived dark mediators, gamma rays can escape the Sun, and neutrinos will be less attenuated. In this work, we have demonstrated gamma-ray and neutrino telescopes are extremely sensitive to such scenarios. Specifically, in this paper we have:

\begin{itemize}
 \item defined a general framework for DM annihilation to long-lived mediators in the Sun,
 \item calculated new solar gamma-ray limits on DM annihilation to long-lived mediators with the Fermi Gamma-ray Space Telescope,
 \item calculated the first solar gamma-ray projections of DM annihilation to long-lived mediators with ground-based water Cherenkov telescopes HAWC and LHAASO, and
 \item calculated new neutrino projections on DM annihilation to long-lived mediators in the Sun with neutrino telescopes such as IceCube and KM3NeT.
\end{itemize}

Experimentally, our results are especially pertinent due to new and upcoming opportunities in the gamma-ray and neutrino channels. For gamma rays, new detailed measurements of the Sun have been made in the GeV range with Fermi, and great increases in sensitivity in the TeV range will be available with HAWC and LHAASO. For neutrinos, the long-lived mediator scenario opens the previously inaccessible multi-TeV window, with gigaton neutrino telescopes such as IceCube and KM3NeT. If the dark sector contains a DM candidate along with a sufficiently long-lived mediator, these telescopes can improve sensitivity to the DM spin-dependent scattering cross section by several orders of magnitude, relative to present searches for high-energy neutrinos from the Sun, as well as direct detection experiments.

Models which have non-suppressed spin-independent scattering cross sections must satisfy strong constraints from direct detection experiments. For models where the spin-dependent cross section is dominant, direct detection limits are significantly weaker. However, if the model contains a long-lived mediator, a substantial part of the spin-dependent scattering cross section can instead be covered via solar gamma rays or neutrinos. Depending on model details, these searches can provide a probe of the spin-dependent DM scattering cross section stronger than the predicted sensitivity for all upcoming direct detection experiments, including DARWIN. This means that observations of solar gamma rays and neutrinos are a promising complementary avenue for the discovery of DM.

Our results define model-independent spaces, from which model-dependent results can be extracted. Demonstrating the optimal model-independent case, as we have in this paper, highlights the maximal power of solar gamma-ray and neutrino signatures. Indeed, further theoretical work is needed to fully explore this parameter space, to interpret it in the context of particular models, and to constrain it with other considerations.  Until then,  the most important thing for progress is new experimental analyses, especially taking advantage of the huge increase in sensitivity possible with present HAWC and LHAASO TeV gamma-ray data, and IceCube neutrino data. Future analyses and accompanying theoretical investigations for long-lived mediators in the Sun have substantial potential to provide crucial insight to the nature of DM.

\section*{Acknowledgments}
We thank Annika Peter, Bei Zhou, Brian Batell, Carsten Rott, Flip Tanedo, Jonathan Feng, Kfir Blum, Maxim Pospelov, Nicholas Rodd and Nicole Bell for helpful comments and discussions. RKL thanks JFB, KCYN, and the Center for Cosmology and AstroParticle Physics (CCAPP) at Ohio State University for their hospitality and support during her visit, where most of this work was completed. RKL was supported by the Australian Research Council and NSF Grant PHY-1404311. KCYN was supported by the Ohio State University Presidential Fellowship and NSF Grant PHY-1404311. JFB is supported by NSF Grant PHY-1404311. Figure~\ref{fig:sun} is drawn using Ti$k$Z~\cite{tantau:2013a}.

\bibliography{longlived.bib}

\end{document}